\documentclass[sigconf]{acmart} 
\AtBeginDocument{%
  }

\copyrightyear{2026}
\acmYear{2026}
\setcopyright{cc}
\setcctype{by}
\acmConference[ASSETS '26]{The 28th International ACM SIGACCESS Conference on Computers and Accessibility}{October 25--28, 2026}{Vila Nova de Gaia, Portugal}
\acmBooktitle{The 28th International ACM SIGACCESS Conference on Computers and Accessibility (ASSETS '26), October 25--28, 2026, Vila Nova de Gaia, Portugal}
\acmDOI{10.1145/3797867.3828973}
\acmISBN{979-8-4007-2521-0/2026/10}

\usepackage[dvipsnames]{xcolor}   
\usepackage{etoolbox}             
\usepackage{ifthen}
\usepackage{xspace}
\usepackage{xstring}
\usepackage{soul}
\usepackage{cleveref}
\usepackage{paralist}
\usepackage{enumitem}
\usepackage{comment}
\usepackage{booktabs}
\usepackage{csvsimple}
\usepackage{siunitx}
\usepackage{graphicx} 
\usepackage{pgfmath}
\usepackage{xfp}
\usepackage{cancel}
\usepackage{listings}
\usepackage{xcolor}
\usepackage{xparse}
\usepackage{csvsimple}
\usepackage{tabularx} 
\usepackage{booktabs}
\usepackage{todonotes}
\usepackage{etoolbox}
\usepackage[normalem]{ulem}
\usepackage{markdown}
\usepackage{array}
\usepackage{longtable}
\usepackage{dblfloatfix}

\newtoggle{comments}
\toggletrue{comments} 

\newcommand{\studyinstrument}{ProgramAT\xspace}

\DeclareRobustCommand{\GK}{
    D1\xspace
}

\DeclareRobustCommand{\AZ}{
    D2\xspace
}

\DeclareRobustCommand{\AJ}{
    D3\xspace
}

\DeclareRobustCommand{\KL}{
    D4\xspace
}

\DeclareRobustCommand{\AS}{
    D5\xspace
}

\newcommand{\Quote}[1]{%
   \emph{``#1''\xspace}
}

\newcommand{\etal}[1]{%
   #1 \emph{et al.}
}

\newcolumntype{Y}{>{\centering\arraybackslash}X}

\begin{document}

\title{Bespoke Visual Assistance: What and How do Blind and Low-Vision People Create with Agentic Programming?}

\author{Ellie Seehorn}
\email{seehorn@umich.edu}
\orcid{0000-0002-6949-9079}
\affiliation{%
  \institution{University of Michigan}
  \city{Ann Arbor}
  \state{Michigan}
  \country{USA}
}

\author{Gene S-H Kim} 
\email{genekim@mit.edu}
\orcid{0000-0001-9514-4610}
\affiliation{%
  \institution{Massachusetts Institute of Technology}
  \city{Cambridge}
  \state{Massachusetts}
  \country{USA}
}
\author{Aziz Zeidieh} 
\email{azeidi2@illinois.edu}
\orcid{0009-0000-9334-8660}
\affiliation{%
  \institution{University of Illinois Urbana-Champaign}
  \city{Champaign}
  \state{Illinois}
  \country{USA}
}
\author{Ather Jammoa} 
\email{atherjammoa@yahoo.com}
\orcid{0009-0006-3016-4761}
\affiliation{%
  \institution{Independent Consultant}
  \city{Detroit}
  \state{Michigan}
  \country{USA}
}
\author{Kun Lee} 
\email{klee746@gatech.edu}
\orcid{0009-0009-6988-5808}
\affiliation{%
  \institution{Georgia Institute of Technology}
  \city{Atlanta}
  \state{Georgia}
  \country{USA}
}
\author{Aditi Shah} 
\email{shah.aditi@microsoft.com}
\orcid{0009-0006-8134-9934}
\affiliation{%
  \institution{Microsoft Corporation}
  \city{Redmond}
  \state{Washington}
  \country{USA}
}

\author{Jaylin Herskovitz}
\orcid{0000-0002-9049-5056}
\email{jaylin.herskovitz@tufts.edu}
\affiliation{%
  \institution{Tufts University}
  \city{Medford}
  \state{Massachusetts}
  \country{USA}
}

\author{Anhong Guo}
\orcid{0000-0002-4447-7818}
\email{anhong@umich.edu}
\affiliation{%
  \institution{University of Michigan}
  \city{Ann Arbor}
  \state{Michigan}
  \country{USA}}
\authornotemark[1]

\author{Venkatesh Potluri}
\orcid{0000-0002-5027-8831}
\email{potluriv@umich.edu}
\affiliation{%
 \institution{University of Michigan}
 \city{Ann Arbor}
 \state{Michigan}
 \country{USA}}
\authornote{Equal supervision.}

\renewcommand{\shortauthors}{Seehorn et al.}

\begin{abstract}
AI-powered assistive technologies have long supported blind and low vision (BLV) people in everyday tasks, but they are general-purpose and often fall short of meeting complex, individualized, in-situ accessibility needs. Though agentic programming tools, like GitHub Copilot, have the potential to bridge this gap by lowering the technical barriers to building personal AT using natural language, the practical applicability of this creation paradigm has been unknown. We address this knowledge gap through a two-phase longitudinal co-design study with five tech-savvy BLV users using \studyinstrument, an agentic programming tool that supports the creation, iteration, and testing of camera-based AT. Overall, co-designers created over 37 tools, with some addressing needs unmet by any existing commercial AT—such as identifying Uber rides or interpreting hand gestures. Qualitative feedback from our co-designers and analysis of development logs surface BLV strategies for tool creation, along with key challenges including model capability limits, specification conflicts, and barriers to successful creation. We discuss recommendations to provide appropriate conversational scaffolding, community tool sharing capabilities, and support for specialized models and personal datasets for future agentic programming environments to empower BLV users to create bespoke visual assistance for themselves.

\end{abstract}



\begin{teaserfigure}
\centering
  \includegraphics[
  width=\textwidth,
  alt = {An overview of the research approach. Part 1: Bespoke assistive technology study instrument: ProgramAT. (A) A mobile interface for creating bespoke AT via request to coding agents (shows an interface with a text prompt box for requesting a new AT tool). (B) Structure to instantiate requests as camera-based AT run from a mobile device (shows a custom tool called playing card detector running on the phone's camera feed, detecting a hand of cards). Part 2: 2 month study deployment timeline. Week 1: Phase I kickoff. Weeks 2-4: Deployment, with 15-min check-in interviews and diary entries. Week 5: Phase I wrap-up. Timeline repeats for the 4 weeks of phase II. Part 3: Research Outcomes. Process: 5 blind expert co-designers, 37 bespoke camera-based assistive tools, 68 tool iterations, 24+ hrs of interview data. Research questions: What bespoke AT would BLV people create? Why do people want to create bespoke AT? What tools/ support are necessary to fully realize goals?}
  ]{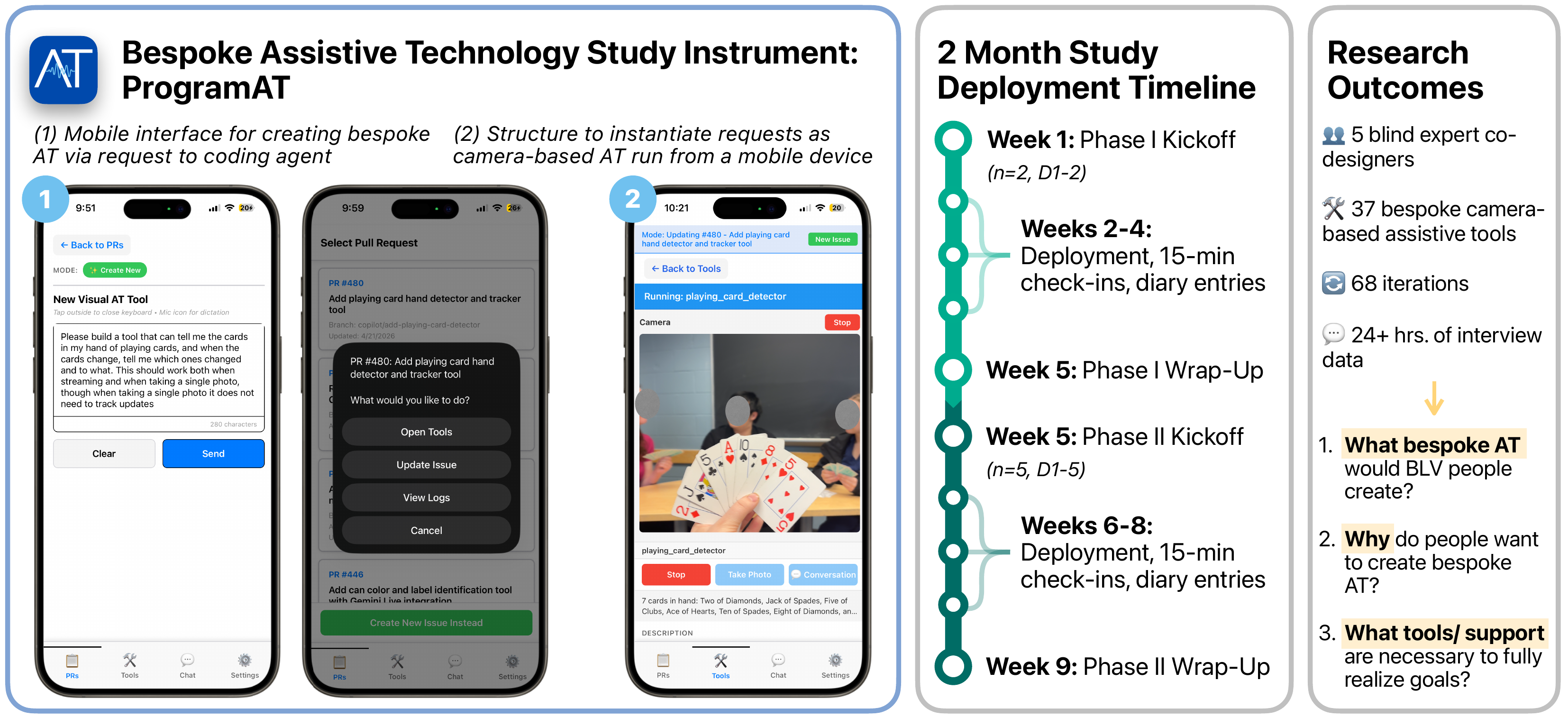}
  \caption{This paper explores how agentic programming might be used to create personal, bespoke assistive technologies. We created a mobile agentic programming interface (ProgramAT) as a means to study this through a longitudinal deployment and co-design. Through a two month study, co-designers created 37 bespoke assistive technologies. We analyze what was created, why, and how we might better support assistive technology creation in the future.}
  \Description{An overview of the research approach. Part 1: Bespoke assistive technology study instrument: ProgramAT. (A) A mobile interface for creating bespoke AT via request to coding agents (shows an interface with a text prompt box for requesting a new AT tool). (B) Structure to instantiate requests as camera-based AT run from a mobile device (shows a custom tool called playing card detector running on the phone's camera feed, detecting a hand of cards). Part 2: 2 month study deployment timeline. Week 1: Phase I kickoff. Weeks 2-4: Deployment, with 15-min check-in interviews and diary entries. Week 5: Phase I wrap-up. Timeline repeats for the 4 weeks of phase II. Part 3: Research Outcomes. Process: 5 blind expert co-designers, 37 bespoke camera-based assistive tools, 68 tool iterations, 24+ hrs of interview data. Research questions: What bespoke AT would BLV people create? Why do people want to create bespoke AT? What tools/ support are necessary to fully realize goals?}
  \label{fig:teaser}
\end{teaserfigure}

\maketitle

\section{Introduction}
AI-powered assistive technologies (AT) such as SeeingAI~\cite{microsoft-seeingAI}, BeMyAI~\cite{bemyai}, and OKO~\cite{oko} have long supported everyday tasks for blind and low-vision (BLV) users\footnote{In this work, we use a mixture of identity-first and person-first disability terminology, to reflect the diversity of disability identity and language preferences across sub-communities.\cite{identity_or_person_first}. We use identity-first language to refer to the BLV community, as our co-designers prefer.}. 
These tools, however are designed to be general purpose and to fit simple, common use cases \cite{hsc}, leaving them insufficient for many complex or individualized accessibility needs. 
For example, these tools do not cater to tasks that involve multiple steps (e.g. checking into a hotel), require invidualized context (e.g. finding personal objects) and related to personal interest (e.g. applying makeup).~\cite{hsc,personal_object_recognizer,lee2020teachable,visualart,applyingmakeup}. 
The one-size-fits-all approach to AT has detrimental consequences to completing desired tasks efficiently \cite{hurst2011empowering, hsc} and with confidence \cite{alharbi_misfitting}, contributing to tool abandonment \cite{phillips1993predictors}. 
Recent work investigates DIY approaches for AT creation, such as structured authoring of formulaic-purpose tools \cite{ProgramAlly}, surfacing augmentative functionality on top of commercial tools \cite{a11yextensions}, or using LLMs as support for physical making processes \cite{kosa2026not}.
These efforts highlight a promising trajectory toward end-user empowerment. However, a gap remains between the desire for customization and the practicality of broad implementation. While many users prefer bespoke AT tailored to their specific needs \cite{bespoke_reflections}, realizing such solutions depends on intensive collaboration with expert designers, making routine personalization impractical.
BLV users, however, have clear ideas of what they would want to accomplish if tools allowed it \cite{gamage2023, hsc}; thus, the core problem is not in users not knowing what they need, but rather, in their lacking accessible means to author AT in-situ.

Agentic coding tools, such as GitHub Copilot, have advanced possibilities by integrating natural language with code generation, thus lowering the technical barrier to building complex systems. This paradigm promises a future where anyone can be a programmer: able to build customized tools and workflows for any use case they desire. For BLV users, this offers the ability to author bespoke AT themselves, in-situ, without needing to wait for a commercial tool to address their challenges, or for their environment to become more accessible.
However, the extent to which such tools enable the creation of useful custom-camera based AT
for BLV users remains unknown. Since BLV users often cannot confirm visual results, the obfuscation of how tools actually operate makes identifying tools appropriate for a desired workflow exceedingly challenging \cite{everyday_uncertainty}. 

To probe into the emerging opportunities and challenges of adopting agentic programming for creating bespoke visual assistance, we seek to answer three guiding research questions:

\textbf{RQ1:
If the barriers to creating bespoke visual assistive technologies were lessened, what would BLV people be interested in creating?}  
By offering a paradigm that widens the range of possibility for what can be created, we aim to achieve a clearer sense for what challenges and desires exist in the long tail of user needs for camera-based AT. In answering this question, we aim to provide insights both into pervasive unmet needs within the BLV community and into the diverse design space required to support more niche, individual desires.
\textbf{RQ2: Why do people want to create bespoke, camera-based assistive technology?} Beyond the specific tools that BLV users may choose to build, we examine the motivations underlying bespoke creation in the camera-based AT domain—including the workflows it enables, and the affordances it provides beyond the resulting tools themselves. By addressing this question, we aim to better understand if and how the value of ``bespoke'' in assistive technology contexts extends beyond the commonly cited potential to address individualized challenges.
\textbf{RQ3: What structures and supports are necessary for end-users to realize the tools they envision using agentic programming?} 
Agentic programming positions itself as offering a wider range of creation possibilities than previous approaches to end-user programming. Yet, AI, both in code authoring contexts and when used in the actual functionality of tools, is imperfect—especially in response to ambiguous natural language. We provide insights into the creation practices people adopt to navigate this precarity, as well as present potential design supports to better empower bespoke creation. 

To investigate these questions, we design \studyinstrument, an agentic programming study instrument that enables BLV users to author bespoke, camera-based assistive technologies by specifying behavior in natural language and iteratively refining executable artifacts, with code and associated edits performed by GitHub Copilot (Figure \ref{fig:teaser}). \studyinstrument supports the integration of vision models, task-specific logic, and contextual prompts, allowing users to create, test, and adapt custom tools in situ.
We use \studyinstrument to approach our research questions through a two-stage co-design process with tech-savvy BLV end-users (two in phase I, five in phase II) over two months. 
In Phase I, two participants engaged in exploratory authoring to provide preliminary insights and inform system refinement to best support intended use cases. In Phase II, five participants created and iterated on multiple custom AT artifacts over time.  Across both phases, participants created a total of 37 camera-based AT prototypes, through a total of 68 iterations. Throughout this process, participants also made per-tool diary entries and participated in weekly check-ins, providing longitudinal insight into the authoring practices and challenges, as well as situations where AI was more likely to succeed or fail.

Overall, this work makes three primary contributions:
\begin{enumerate}[topsep=0pt]
    \item An open source study instrument, \studyinstrument
    \footnote{
    ProgramAT is open-sourced at:\href{https://github.com/program-at/ProgramAT-opensource}{github.com/program-at/ProgramAT-opensource} 
    }
    , that enables the creation, iteration, and testing of bespoke camera-based assistive tools through agentic programming;
    \item Understanding of the types of tools that BLV people may want to create, contextualized with their motivations and strategies to creating these tools; and
    \item Recommendations to design agentic coding environments to support conversational programming of AT.
\end{enumerate}

\section{Related work}
While commercially available tools are are widely used, they leave several gaps, particularly with respect to nuanced, individual use cases. We make these gaps explicit and discuss open problems in the area in AI in visual assistive technology (\Cref{subsec:AI_VAT}).
Addressing such gaps in accessible, personalized assistive technology demands approaches that simultaneously support customization, expand participation in creation, and foreground disabled perspectives. We position our work within this intersection, drawing on research in bespoke assistive technology (\Cref{subsec:bespoke_at}),  end-user programming (\Cref{subsec:end-user-programming}), and disability-led design (\Cref{subsec:disability-led-design}).

\subsection{AI-powered Visual Assistive Technology}
\label{subsec:AI_VAT}
Commercially available AI-powered AT (e.g. SeeingAI~\cite{microsoft-seeingAI}, BeMyAI~\cite{bemyai}, OKO~\cite{oko}) receive widespread use within the BLV community; however, recent research documents that for many complex or individual needs, these tools are insufficient \cite{hsc}.

One area of insufficiency surrounds tools that require multiple steps: often requiring frequent tool switches to achieve a multi-step goal \cite{hsc}.  
Commercial AI tools, even when successful in supporting one step of a user's goal, routinely fail to support them throughout the course of the interaction \cite{beyond_visual_perception}: failing to suggest or consider proactive next steps in favor of treating only the moment at hand. Further, across multi-step interactions, assistive AI regularly experiences task-drift—reducing its guidance's efficacy and increasing cognitive load
\cite{zhao2025less}.

Another major area in which commercially available AI-AT faces challenges is in tasks related to personal interests. 
Without knowing a user's personal standards or expectations, generalist tools cannot provide reliable, adequate, contextualized guidance. This leads to answers that are inconsistent on a single input (e.g. reporting if ``enough'' of a makeup product has been applied ~\cite{makeup_steps}) or too general for use in a precise context (e.g. describing the piece needed to complete a puzzle section only as ``square'' \cite{beyond_visual_perception}).
This failure is not unique to assistive AI: general purpose AI, like ChatGPT's voice mode, also tends to fall back on general, non-situationally grounded knowledge when asked about niche or individual scenarios \cite{probing_the_gaps}. 
Further, in contexts where immersion and interpretation, rather than task-completion, are the aim, the goal-driven model of assistance falls short.  This challenge presents barriers to using commercial AI AT in situations where there is not one "correct" answer (e.g. art interpretation \cite{visualart}), or the desired answer is informed by the user's own perspective and context (e.g. taking in one's surroundings while exploring nature \cite{tang_nature}).

A third area where commercial AI AT historically struggles is in cases requiring individual context, such as finding personal objects \cite{personal_object_recognizer, lee2020teachable}. Recent commercial AI AT has made progress in this area: Find My Things \cite{wen2024find} has recently been integrated as a SeeingAI mode \cite{microsoft-seeingAI}, empowering users to leverage generalist AT towards individual purposes in some cases.

However, advances to personalizing widespread AI-based AT come with new privacy and security risks. Users have little control over what models are used and what data is received and retained in commercial tools. Prior work surfaces an expectation of limited data retention that by tools that require permanent upload of data to achieve personalization cannot meet \cite{stangl2023dump}, and highlights anxieties around using AI AT in private settings for fear of capturing sensitive content \cite{emerging_obfuscation, before_i_asked_my_mom}. Recent work supports obfuscating private information to lessen these concerns \cite{designing_obfuscation}. Yet, BLV users cannot readily verify obfuscation results \cite{trying_to_piece_it_together}: limiting obfuscation alone as risk-mitigation strategy in privacy-aware AI AT. 

These concerns are relevant to any use of AI in visual assistance, not just generalist, commercially available tools. Still, one major design guideline for privacy-protecting GenAI use, as proposed in 
\etal{Sharma}
\cite{before_i_asked_my_mom}, is explicit user consent when data processing occurs over-network or off-device. While many AT may require the use of off-device models, users can more readily give informed consent when the models in use are 1) transparent and 2) substitutable such that if a user does not consent to a particular model or service, they can change the model rather than disqualifying use of the tool as a whole.

\subsection{Bespoke Assistive Technology}
\label{subsec:bespoke_at}
In response to the barriers presented by general-purpose AT, disabled users—both within and beyond the BLV community—often attempt to combine \cite{hsc} and augment existing tools \cite{irobot, a11yextensions}, or develop narrowly-scoped custom tools \cite{ProgramAlly}.
Still, these solutions are labor intensive, and as a product of being retrofit rather than a native solution, often still have gaps that prevent fully addressing accessibility needs \cite{wu2021cripping}.
The usability problems that follow lead to many being stuck with AT that is inadequate for their needs, or having a long list of ATs that have been tried, most of them abandoned \cite{hurst2011empowering}. Thus, many end-users prefer bespoke AT \cite{bespoke_reflections}: expert-built tools tailored to their needs. 
However, making such systems demands close collaboration with designers, rendering everyday customization impractical. 

Supports to build bespoke AT vary in the types of creation and customizations that are supported. Historically, bespoke AT manifests in the physical domain \cite{a11ybits, sharingIsCaring}, driven by the the maker movement \cite{makermovement}. Clinical professionals \cite{aflatoony2023collective, slegers2020makers, junk_brilliant} and disabled end-users \cite{thorsen2021patient, comingToGrips, aflatoony2023collective} alike have embraced digital fabrication as a means to produce custom assistive devices \cite{sharingIsCaring, sarwargrassroots}. 

Another, emergent, line of bespoke AT research leverages generative AI to enable custom accessibility experiences across a range of applications \cite{kosa2026not}.
One application is augmentation, where GenAI supports goals beyond basic access for users of existing AT such as enabling social integration through humor and personal voice \cite{whysoserious, irobot, chatGPTasAT} in use of AACs—tools that support non-speech communication. Other applications enable the creation of entirely new experiences, like natively nonvisual online shopping for BLV users \cite{echomall} and rapid prototyping interfaces that allow occupational therapists to build tools that meet diverse access needs \cite{LiGenAIDIY}. 
Generative AI is uniquely powerful in the context of bespoke AT, as it has rapidly expanded the range of assistive tasks that are perceived as plausible to address with technology \cite{choi2025farillgo}. In this work, we leverage this broadened possibility space to gain greater insights into the true interests BLV users have in creating bespoke AT. 

\subsection{End-User Programming}
\label{subsec:end-user-programming}
In the software domain, the majority of bespoke AT creation falls into one of two categories. The first is hyper-personalization, generally done by very technically savvy disabled people by standard development methods \cite{irobot, DHHtraveler}. The other—which this work focuses on—is end-user programming, defined by \etal{Ko}. as programming done by people who may or may not have technical expertise in order ``to support some goal in their own domains of expertise’, further, ‘to achieve the result of a program primarily for
personal, rather [than] public use'' \cite{koenduser}. Though used by people of varied levels of programming skill, end user programming generally tries to streamline the programming experience \cite{wong2007making} to lower the floor (reduce barriers to entry \cite{resnick2009scratch}), raise the ceiling (expand what is possible \cite{matsuzawa2015language}), or both. 
Our work deepens understanding along this line of work, aiming to both lower the floor and raise the ceiling by making end-user programming more accessible while expanding the range and complexity of assistive tools that BLV users can create, relative to prior work\cite{ProgramAlly}.

End-user programming for accessibility takes many forms \cite{herskovitz2024diy}. Examples include simple block based programming for accessible classrooms \cite{r2025blocks4all}, a templated domain-specific language for finding specific visual information \cite{ProgramAlly}, leveraging on-device automations to surface novel, relevant functionality \cite{a11yextensions}, and customizable AI interfaces used by people with disabilities \cite{leeDIYaphasia} and medical practitioners \cite{zastudilProgrammingAAC} alike for just in time adaptive communication. 
The advent of agentic programming opens up new possibilities for conversational programming, a type of end-user programming. Conversational programming predates LLMs \cite{chilana2015perceptions}, yet, the prevalence of AI coding agents, like Copilot \cite{copilot}, Cursor \cite{anysphere}, and Claude Code \cite{claudecode}, reveal the possibility for this natural language form of programming to be usable by the general populace \cite{akhoroz2025conversational}. These agents have already been used in several accessibility contexts: from web development \cite{mowar2025codea11y} to sound-accessible AR \cite{lee2025sonocraftar}. 

Though code agents have begun to be applied in accessibility contexts, it is still not well understood how they support end-to-end creation of bespoke assistive tools in practice. This work addresses this gap by examining how technically experienced BLV users create camera based assistive tools with agentic programming systems, focusing on what they create, how they navigate the development process, and where their processes succeed or breaks down.

\subsection{Disability-Led Design}
\label{subsec:disability-led-design}
The existence of conversational programming as an approach to authoring bespoke AT does not, by itself, make it a suitable solution for addressing long-tail accessibility needs.
Indeed, recent work indicates that the types of problems visual accessibility research often tackles are misaligned with the problems the BLV community places value in solving \cite{gamage2023}.

Disabled users are not waiting idle for research to meet their needs, but often trailblaze innovation based in lived experience in their own right \cite{aashakainnovator, earlyadopter}. By including voices of these innovators explicitly in the research process and allowing lived experience to drive design decisions, the tacit knowledge of lived experience can be translated into material scientific knowledge.

Such an approach is critical in accessibility contexts, where needs are shaped not just by functional requirements, but by community-specific practices. Prior work shows that disabled communities often develop distinct cultural norms and interaction practices, such as differences in communication styles among AAC users \cite{backchannel} and linguistic and interactional preferences within the DHH community \cite{keating2003american}. At the same time, accessibility needs are highly context-dependent, shifting across environments and social situations\cite{shuxuDIYdeaftech, wheelchair_control}. Questions of discretion and self-preservation further shape how and when assistive technologies are used \cite{curtisAAC}. Co-design helps surface these cultural and contextual factors, supporting the development of assistive tools that better align with users' lived experiences, priorities, and forms of self-representation \cite{DIDwearable}.

To simultaneously address the need for community participation and faithfully answer the target research questions, we engage in the established strategy of involving working prototypes in participatory design \cite{haworth2016use}. This practice is standard in technically driven co-design in accessibility \cite{a11yextensions, collabally, jeremy_designforone, dishdetect} and HCI more broadly \cite{mrtree, namubot, enhancingcraftman}. We choose co-design here because the lead author is not part of the BLV community: this method ensures the design is driven by community perspectives.

\section{Study Instrument: \studyinstrument}

To support answering our research questions, we developed \studyinstrument: a study instrument that allows users to engage in natural-language creation, refinement, and use of bespoke camera-based assistive tools. It consists of a mobile app, an agentic coding environment (GitHub Copilot retrofitted with scaffolds and tools), and a compute server. With \studyinstrument, users can prompt for the creation of a desired tool using natural language—a format demonstrated to be highly expressive \cite{krings2025r}. Created tools use the camera as input and output speech, audio tones, haptics, or some combination thereof at the user's specification.  \studyinstrument differs from existing AT apps (e.g. BeMyAI) through its ability to generate, refine, and execute task-specific code rather than relying on predefined functionality or thin LLM calls. These functionalities broaden the opportunities for what bespoke AT could accomplish, bolstering our ability to answer the question of what and how BLV people want to create. 

\subsection{Usage Walkthrough}

\begin{figure*}[t]
    \centering \includegraphics[
    width=\linewidth,
    alt = {A walkthrough of the mobile ProgramAT creation interface. In development mode: (1) Initial tool creation prompt (shows a text box with a written request for a car identification tool), (2) Programming agent processes request (mobile requests are sent off-device to the copilot agent in personal repository, copilot agent implements request following programAT system prompt guidelines ensuring modularity), (3) Test, debug, find areas for change (a screenshot of the car identifier tool running, the camera view shows an image of a car pulled up on a laptop screen for easy testing), (4) Request iterations (another text box with a request to make the car descriptions shorter, commenting on the open pull request made by the agent). In production mode: (1) Choose tool to run from list (a list of available tools including car identifier and scene description), (2) Run with camera view (shows the tool running on a picture of a car in a parking lot). Production mode shows tools that have been merged to main (and whose components can be reused in future tools), and also hides pull requests for a simpler interface.}
    ]{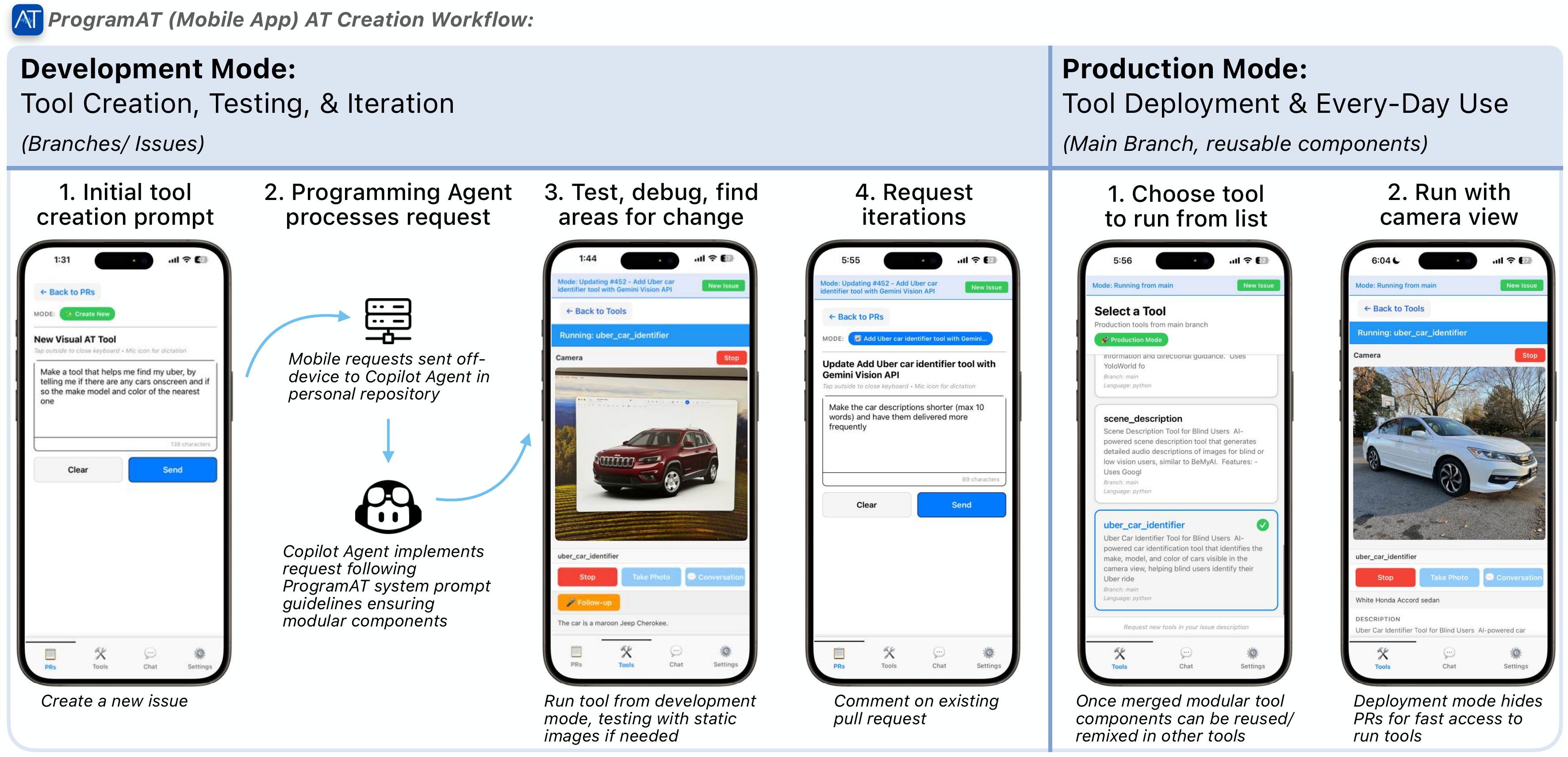}
    \caption{A walkthrough of how users create and use AT tools through the ProgramAT mobile app interface. In \textbf{deployment mode}, users create, test, and iterate on tools. (1) A request for a new tool is made, opening a GitHub Issue. (2) The programming agent processes the issue, following ProgramAT structure to create modular tools that run in a mobile camera-based environment. (3) Users test the created tool using the camera, observing issues. (4) Users request iterations to the tool by commenting on the existing pull request. Then, in \textbf{production mode}, users access a simplified interface for every day use, hiding pull requests and incomplete tools. They (1) choose the tool they want to run from a list, and (2) run it from the camera view.}
    \Description{A walkthrough of the mobile ProgramAT creation interface. In development mode: (1) Initial tool creation prompt (shows a text box with a written request for a car identification tool), (2) Programming agent processes request (mobile requests are sent off-device to the copilot agent in personal repository, copilot agent implements request following programAT system prompt guidelines ensuring modularity), (3) Test, debug, find areas for change (a screenshot of the car identifier tool running, the camera view shows an image of a car pulled up on a laptop screen for easy testing), (4) Request iterations (another text box with a request to make the car descriptions shorter, commenting on the open pull request made by the agent). In production mode: (1) Choose tool to run from list (a list of available tools including car identifier and scene description), (2) Run with camera view (shows the tool running on a picture of a car in a parking lot). Production mode shows tools that have been merged to main (and whose components can be reused in future tools), and also hides pull requests for a simpler interface.}
    \label{fig:sys}
\end{figure*}
We demonstrate the \studyinstrument's creation and iteration capabilities through a walkthrough of creating an Uber finder tool (a use case for an unmet need expressed by our co-designers) (shown in Figure \ref{fig:sys}). To create such a tool, a user starts by prompting for the tool, either through the mobile app using natural language or through the GitHub site, using the `visual assistive technology' template. If using natural language, the user's prompt is mapped to the fields of the template, and if prompting from the web into the template, the user's prompt is used directly. This request is then assigned to GitHub Copilot, a coding agent that proceeds to implement the user's requested tool, following the guidelines described in Appendix \ref{appendix:instructions} 
Once generation is complete, the user runs the tool from the mobile app, pointing the camera at the street and receiving real-time feedback about the presence of cars and potential Ubers. If they want to make any changes to the tool, for instance, reducing the verbosity of responses, they may do so in the mobile app by choosing to update issue, or on the GitHub site by leaving a comment on the pull request and tagging Copilot. This prompts a code update, which the user can resume testing once implemented. Once satisfied with the tool, the user visits the GitHub site to merge the pull request—transitioning the tool from development to production. On production, the tool is runnable for everyday use and its components can inform the implementation of future tools.

\subsection{Key Interaction Components}
\subsubsection{Modular building blocks}
We instruct Copilot to generate code in a modular, building-block style, where each tool is composed of reusable components that encapsulate specific functionality. Rather than directly modifying existing tools, \studyinstrument instructs Copilot to compose existing tools' components and augment them with new code only when none yet exists to satisfy a particular purpose, enabling tools to be developed incrementally without introducing dependencies on prior implementations \cite{zaman2025whatsai}. Individual tools remain self-contained while still integrating with shared system infrastructure. As a result, users can build increasingly complex tools by composing modular building blocks across successive pull requests. 

\subsubsection{Intelligent model selection}
\studyinstrument selects models based on task type rather than applying a single model uniformly. Object detection defaults to YOLO11 (COCO dataset), with YOLOWorld used for objects outside COCO's class set. Text extraction uses the Google Cloud Vision API. General reasoning and vision-language tasks use Gemini models. When a user requests a Custom-GPT-esque \cite{customgpt} interaction, \studyinstrument uses Gemini Live and automatically reissues queries at regular intervals, overcoming the cumbersome reprompting present in standard video-AI interactions~\cite{probing_the_gaps}.

\subsubsection{Mobile and desktop integration}
ProgramAT's agentic coding environment can be accessed from the mobile app, or directly from the GitHub web desktop interface, where users can create tools by creating standard GitHub issues \cite{githubissue_2025} and iterate by commenting on pull requests. An example of the tool creation process from the web is shown in Appendix \ref{appendix:githubweb}, Figure \ref{fig:web}. This flexibility can support a variety of creation workflows.

\subsubsection{Iterative workflow}
After testing a tool, users can iterate by submitting a natural-language update request from the app or leaving a comment on the GitHub pull request on the web. Copilot applies changes under the same architectural constraints as the original generation, and users can re-run and re-evaluate tools as soon as re-generation concludes, supporting a tight loop between real-world use and tool refinement.

\subsubsection{Development and production modes}
\studyinstrument distinguishes between \emph{Development Mode}—where users create and refine tools via pull requests—and \emph{Production Mode}, where finalized, merged tools are available for streamlined use. This separation lets users freely experiment without disrupting tools they already rely on.

\subsection{User Interface}
The \studyinstrument mobile app is organized around two primary views corresponding to its two modes. In \emph{Development Mode}, the \emph{PRs} tab lists all tools currently under development. Each entry displays the tool's name and current status (e.g., ``WIP'' while the agent is generating code). Users can tap a pull request to select from tools on the associated branch, submit an iteration request, or view associated logs. Selecting a tool on the branch launches it, transitioning the user into a camera interface where the tool can be run. This camera interface supports two interaction modes. In \emph{Stream} interactions, the tool continuously processes live camera frames and delivers outputs—spoken aloud, as haptic feedback, or as on-screen text—as the scene evolves. This is suited for tasks requiring continuous awareness, such as tracking moving objects or monitoring a changing environment. If the tool is built to work similarly to a Custom GPT—a customized version of an off-the-shelf LLM designed to handle a specific task or query \cite{customgpt}—, this mode also allows the user to ask live follow up questions as the tool runs, a \emph{Live} interaction style within the standard \emph{Stream} way of running tools. By contrast, in \emph{Take Photo} interactions, the user captures a single image and receives a response tied to that moment, better suited to one-off queries. After a photo capture, users can ask natural-language follow-up questions; the visual context is retained, so users can progressively build understanding without recapturing input. In all modes, images are discarded from the server post-processing to protect users' privacy.
 

\subsection{System Implementation}
\studyinstrument is implemented in React Native and deployed via iOS TestFlight. Each participant is assigned a personal GitHub repository forked from a shared base, providing an isolated development environment. Tool creation is mediated by a GitHub issue template tailored for camera-based assistive technology, which structures proposals around problem descriptions, example usage, and implementation details—reducing ambiguity and ensuring agent-generated code remains grounded in the user's professed needs. When a user submits a prompt from within the app, it is first parsed by Gemini-3-Flash-Preview to populate the template fields; if any required fields cannot be inferred, the app prompts the user for clarification before submitting. This structured handoff reduces underspecified requests reaching the coding agent.
 
The coding agent (GitHub Copilot) operates under a detailed instruction file (see \Cref{appendix:instructions}) that enforces the modular architecture and model selection logic described above, as well as a set of starter tools. All generated tools conform to a standardized JSON output schema, which the mobile client parses to route outputs to the appropriate feedback channel—speech, earcons, haptics, or on-screen text. This shared schema makes multimodal output possible without tool-specific client logic.
 
A backend server, written in Python, coordinates agent sessions, manages GitHub API calls, handles tool execution by forwarding camera frames to the appropriate model endpoint, and returns structured results to the app. It also surfaces lightweight agent progress updates to the mobile client via WebSocket.

\section{Co-Design Study Method}
Our study's objectives are two-fold: first, to understand \emph{what} visual assistive tools people who are blind or low-vision build, and second, \emph{how} BLV users may approach building these tools. 
We utilize co-design—defined in \etal{Ellis} as "a subset of participatory design which is used to understand individual usage scenarios" \cite{bespoke_reflections}—to explore these objectives.
The study involved five tech-savvy co-designers (two in phase I and three additional co-designers in phase II) building custom visual tools with \studyinstrument, making diary entries, and participating in  semi-structured interviews over a span of two months. Phase I (n=2) was a smaller, exploratory study aimed at generating preliminary insights into what BLV users wanted to create and how they approached creation, while also informing iterative refinements to \studyinstrument to help meet these goals.  Phase II built on the refinements and initial findings as a larger study (n=5) conducted with a more mature, stable version of \studyinstrument, enabling a deeper examination of users’ creation intent and the extent to which AI can meaningfully support the development of bespoke camera-based AT. Each phase lasted approximately one month and consisted of a kick-off meeting, three weekly check ins, and a wrap-up meeting, with each co-designer building and iterating on at least four tools with \studyinstrument throughout the phase. By the end of phase II, our co-designers built a total of 37 tools, with 68 total iterations. 

To answer our research questions, we analyze qualitative data from interviews and diary entries, and conduct AI-assisted code reviews on the code for these tools. 

\begin{figure*}[t]
    \centering \includegraphics[width=\linewidth,
    alt={
    Timeline of the study methods. Research Questions: What bespoke AT would BLV people create? Why do people want to create bespoke AT? What tools/ supports are necessary to fully realize goals? Phase I (1 month, 2 participants D1 and D2): Week 1 Kickoff Session (90 mins, background interview, study instrument onboarding, tutorial task), Week 2-4 Deployment and Check-Ins (Three 15-minute check-in meetings to report progress, diary entries for each tool created via GitHub Issue, minimum of 1 per week), Week 5 Wrap Up Session (90 mins, final 15 minute check in, direct observation of 1 new tool creation, reflection interview on Phase 1 overall). Phase II follows a similar timeline pattern from week 5 to 9, except only Participants D3, D4, and D5 completed the kickoff session, and the wrap up session only included a reflection interview and no direct observation.
    }
    ]{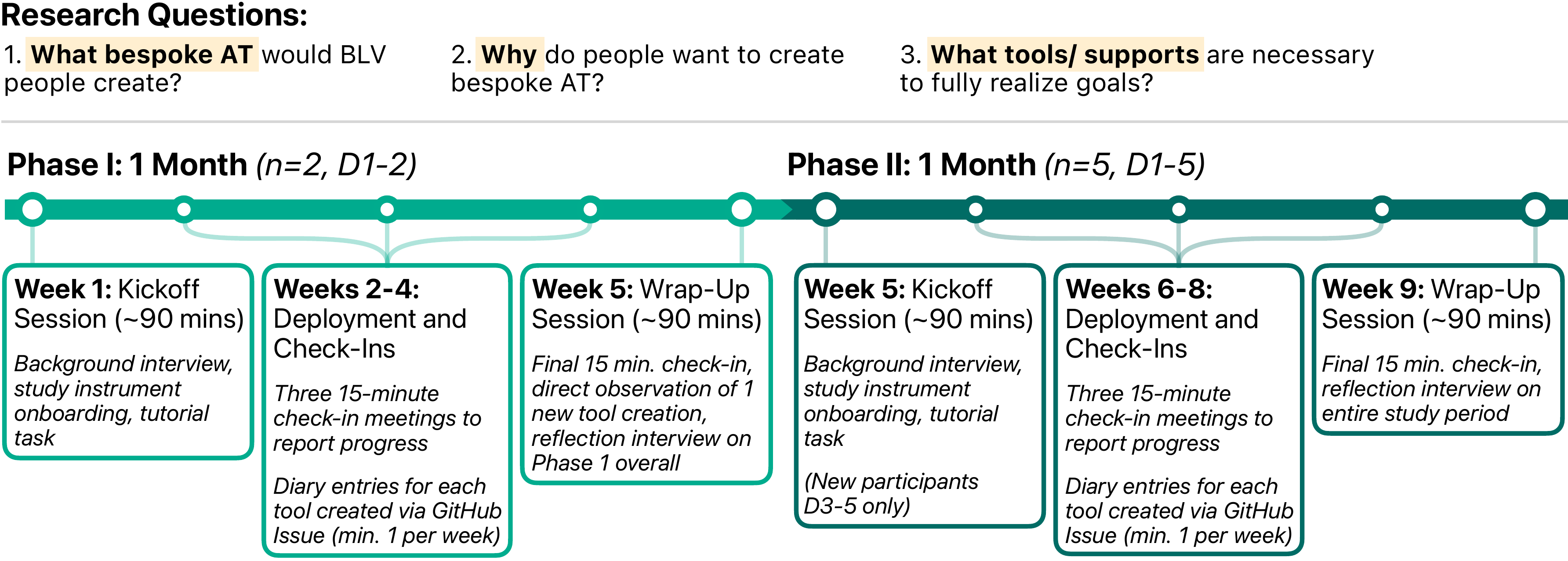}
    \caption{Timeline of the study methods. Phase I: A preliminary deployment with two co-designers to refine the study instrument. Phase II: A deployment of the final study instrument with five co-designers for further data collection. Each phase included a kick-off interview with a system tutorial, a four-week deployment with three 15-minute progress check-in meetings and diary entries, and a wrap-up reflection interview.}
    \Description{Timeline of the study methods. Research Questions: What bespoke AT would BLV people create? Why do people want to create bespoke AT? What tools/ supports are necessary to fully realize goals? Phase I (1 month, 2 participants \D1 and \D2): Week 1 Kickoff Session (90 mins, background interview, study instrument onboarding, tutorial task), Week 2-4 Deployment and Check-Ins (Three 15-minute check-in meetings to report progress, diary entries for each tool created via GitHub Issue, minimum of 1 per week), Week 5 Wrap Up Session (90 mins, final 15 minute check in, direct observation of 1 new tool creation, reflection interview on Phase 1 overall). Phase II follows a similar timeline pattern from week 5 to 9, except only Participants \D3, \D4, and \D5 completed the kickoff session, and the wrap up session only included a reflection interview and no direct observation.}
    \label{fig:study-timeline}
\end{figure*}

\subsection{Co-Designers}
\label{subsec:codesigners}
Across both phases, we engaged with five co-designers, all of whom are authors on this paper. 
Two co-designers (\GK, \AZ) were involved in both phases, while the remaining three co-designers (\AJ, \KL, \AS) engaged with phase II only.
General demographic information for all co-designers is described in Appendix \ref{appendix:demographic_table}, \Cref{tab:participant_demographic}.

Our use of the term ``co-designer'' refers to 1) participants’ collaboration with the research team in shaping \studyinstrument (the primary focus of Phase I, where the individual usage scenario is the creation process), and 2) use of \studyinstrument to make custom tools (the primary focus of Phase II, where the individual usage scenario is the visual access need a tool addresses). \cite{bespoke_reflections} This relationship is both human-human and human-AI. 

Our participant selection strategy leverages the strong influence of technical experience beyond AI literacy alone on effective AI use while being grounded in the lived experiences of people who may need visual assistive tools, enabling exploration of the upper bounds of AI capabilities  to create bespoke visual AT \cite{ma2026not,2023johnny}. Essentially, working with technical users helps distinguish fundamental AI limitations from issues of inexperience \cite{ma2026not}so we adopt the established practice of recruiting experts when relevant to the research questions \cite{three_modalities, a11yextensions}. We recruited BLV co-designers who were tech-savvy, in particular with regards to AI and basic development experience. Technical experience was materially necessary for this study, since effective \studyinstrument use requires comfort with development flows (e.g. Git). Each co-designer's self-reported fluency and confidence across several relevant technical areas is described in \Cref{tab:participant_skill}.

Our focus on tech-savvy BLV co-designers reflects observations that users with disabilities often also represent early adopters of AI based technology \cite{earlyadopter}. \GK and \AZ, both graduate students in computing and information sciences, were recruited for the first phase of our co-design study. Their overlapping experiences as researchers and visual assistive technology users yielded valuable \emph{cripepistemologies}, defined in \etal{Desai} as the way ``innovators'' lived experiences and the corresponding crip knowledge shape both the high-level idea behind their innovations and the smallest of design choices'' \cite{aashakainnovator}.  Their cripepistemologies provided simultaneous familiarity with early prototypes and intuition for the types of AT affordances that best support BLV creation and agency: a relevant combination of experiences for phase I's goal of informing system iteration. For phase II, we added three additional co-designers, with similar technical experience but working in non-research careers to improve population coverage. We recruited these co-designers through personal networks and snowball sampling.

\begin{table*}[h]
    \centering
    \renewcommand{\arraystretch}{1}
    \begin{tabular}{l|l|l|p{0.12\linewidth}|p{0.12\linewidth}|p{0.12\linewidth}}
        ID & Profession & General Tech Literacy & Off-the-Shelf AI &  Development Flows (ex: Git) & Code Agents\\ \hline\hline
        \GK & CS PhD student &7 & 6 & 5 & 5  \\
        \AZ & Informatics PhD student &6 & 5 & 6 & 5  \\
        \AJ & AT Instructor &7 & 7 & 4 & 5 \\
        \KL & Accessibility Engineer &6 & 7 & 5 & 6 \\
        \AS & Software Engineer &7 & 7 & 5 & 6  \\
    \end{tabular}
    \vspace{0.5pc}
    \caption{Co-designer experience and self-reported technical skills (1-7), where 1 is totally novice and 7 is totally expert.}
    \label{tab:participant_skill}
    \vspace{-1.5pc}
\end{table*}

\subsection{Semi-Structured Interviews}
\label{subsec:interview}
Across both phases, our semi-structured interviews onboarded co-designers, and gathered insights when wrapping up phases of the study. These interviews approximately lasted between 90 minutes to 2 hours, were audio-video recorded, and transcribed using the automated transcription features of the videoconferencing platform. The lead researcher memoed after each interview, and corrected transcription errors by hand during analysis. Protocols are available in Supplementary Material.

The onboarding interview captured our co-designers' technical expertise, along with details about their use of AI for accessibility. We then introduced \studyinstrument by asking our co-designers to build an empty seat finder and clothing recognizer—pre-determined tools that address overlooked needs \cite{gamage2023, clothing2017, clothing2018} and ensure consistent tool behavior. Co-designers then familiarized themselves with \studyinstrument by creating a camera-based tool of their choosing, testing and iterating on it.

The phase I wrap-up interview asked co-designers to reflect on the previous four weeks, with questions capturing where they had found success and friction so far, and how those experiences compared to those they had with pre-built AI AT \cite{oko, bemyai, microsoft-seeingAI}. 
This culminated with co-designers creating a tool of their choosing to facilitate observation of their creation and iteration strategies.
Finally, the phase II wrap-up interview reflected on these strategies longitudinally, and highlighted co-designers' perspectives on the future of creating and using bespoke AT.

\subsection{Weekly Check-ins and Diary Entries}
\label{subsec:checkin}
After the onboarding interview, co-designers incorporated \studyinstrument into their daily lives, using it to create tools for their own purposes, with the expectation of producing at least four tools over the four-week period of each phase. To understand their motivation and approach to tool creation, we conducted weekly 15-minute check-ins where we asked a standardized set of questions, exploring what participants had built or iterated on, what was working, what was not, and—particularly for Phase I—how \studyinstrument could better support their goals.

Check-ins provided insight into both the barriers co-designers faced in creating useful, performant tools and the approaches that led to success. They also supported collaborative ideation with the research team and iteration: giving co-designers an opportunity to better understand how \studyinstrument operates in order to guide iteration and future creation, while offering the research team clearer visibility into key pain points and how they might be addressed.

In addition to weekly check-ins, when a co-designer finishes work on a tool—whether they are satisfied to call it complete and merge it to main, or they have devoted substantial effort but remain unsatisfied and elect to abandon it—they “close” the associated pull request by leaving a diary entry as a comment on the PR. This comment describes their idea and prompting strategy, any iterative steps taken after the initial attempt, unexpected system behavior (positive or negative), and how the tool performed. Notably, to the authors' knowledge, using GitHub comments as diary entries may offer a novel and promising method for coding-based qualitative studies; we will explore this further in future work.

Where check-in meetings provide in-the-moment insights into tool performance and the co-designer’s creation process, diary entries offer a post-hoc reflection on the tool-building experience. This reflection enables clearer distinction between barriers that were surmountable through iteration and those that persisted as salient long-term challenges, supporting a more precise thematic understanding of limitations and opportunities in tool creation.

\subsection{Data Collection and Analysis}
Through the two-phase study, we collected interview data, diary entries, and development activity including original prompts, pull request comments that informed iterations, and the code for each tool that our co-designers built.

The onboarding, phase two kickoff, and wrap up interview data was analyzed by two authors generally following the six phases of thematic analysis \cite{braun2006using}. We used thematic analysis as it best fits our research questions; it is not only useful for understanding themes in participants’ behaviors and perspectives \cite{braun2021can}, but also widely recognized as fitting for applying qualitative insights across multiple forms of data \cite{dejonckheere2024qualitative, dwyer2020analysis, fona2023qualitative}.
Two members of the research team (a lead and supporting author) performed the analysis. The supporting author performed the first three phases of analysis on a portion of the 90-minute sessions, generating memos \cite{birks2008memoing}, initial codes, and potential themes. The lead author compared this with their own analysis of the full data to ensure agreement and coverage. The lead author completed the analysis, with the supporting author reviewing the final themes and report.
Additionally, the lead author coded the 15-minute check-in data in conjunction with the final tool diary entries. 

These were analyzed in tandem because of their overlapping purpose: weekly check-ins were intended to provide insight into the intention and experiences behind building tools as they were being built, whereas a diary entry collects the same information post-hoc, providing a higher-level overall view. By considering these pieces together, we gain a more comprehensive view of the creation experience.

Notably, we also conduct a post-hoc code review, assessing the code created at each step, for each tool, by the end of the study.
Following the standard code review practice of tailoring to organization-specific requirements \cite{no_standard_review}, we develop spec-driven code review criteria \cite{piskala2026spec}. To establish specs, we reference the agent instruction file (available in Appendix \ref{appendix:instructions}), the instructions from the co-designer's initial prompt, and, in the case of iterations, any additional or changed instructions provided by co-designers in follow ups. Follow up instructions are included alongside co-designer's initial requests because conversational programming often operates under \emph{progressive specification} \cite{fawzy2025vibecodingpracticemotivations, tang2026programming}, a practice where developers iteratively refine outputs rather than specifying complete tasks upfront \cite{tang2026programming}. The resulting review focuses on whether generated code satisfies the provided specs.

Though agentic programming tools support AI code reviews, there are often gaps in their understanding of user's intent \cite{haider2026understandingdominantthemesreviewing} which lead to poor comprehension of code's meaning in context \cite{chowdhury2024ai}.
However, AI's code-review performance improves when provided example human reviews \cite{lin2024improving}. Thus, we adopt a human-AI approach to our code reviews. The lead author manually reviewed the commits associated with the first Copilot session for each tool as a group—representing Copilot's first attempt at meeting specs—and provide this review along with the earlier established spec-driven guidance to facilitate the AI code review for a tool's remaining Copilot sessions. To account for model diversity, we use Claude Sonnet 4.6 for these AI code reviews: Claude is not used by default in Copilot's generation, and our tools use Gemini when a vision-language model is needed.

Finally, we analyze both initial and iterative prompts alongside the results of code reviews to identify patterns underlying AI successes and failures in tool creation, with the goal of informing future structural supports that better enable BLV end users to author bespoke camera-based assistive technologies. We consider several metrics of ``prompt quality'' as established in \etal{Torka} \cite{Torka2024OptimizingAC}, including general features such as length, presence of example usage, provision of situational context, specification of technical detail, and, in the case of iteration prompts, distance from previous output to desired new output. Further, \studyinstrument offered both mobile and desktop methods of input; we also consider which input modality was used, in order to judge the tradeoffs between the expressiveness mobile's pure natural language interface is theorized to offer \cite{krings2025r} and the precision afforded by a more structured, templated form of input offered in the desktop experience \cite{adepu2016comparison}.

\section{Findings}
Our research questions seek to understand what BLV people would create if barriers to bespoke visual AT were lessened, why they are motivated to create bespoke camera-based AT, and what supports are necessary to realize those tools through agentic programming. To address these questions, we conducted co-design activities with five BLV co-designers who built tools using \studyinstrument over the course of two months. We highlight findings from these activities: we first present the range of tools co-designers envisioned and built—spanning common accessibility needs to highly individualized desires—and examine the motivations underlying their creation choices (RQ1, RQ2); we then turn to the creation process itself, surfacing the strategies co-designers developed to navigate the inherent ambiguity of agentic programming and the design supports that helped them realize their goals (RQ3).

\subsection{What Motivates Tool Creation?}
\textbf{Co-designers built a broad range of bespoke assistive tools, initially prioritizing everyday accessibility needs before progressively exploring more personal, speculative applications as they became comfortable with \studyinstrument.}
\begin{table*}[h!]
    \centering
    \renewcommand{\arraystretch}{1.15}
    \begin{tabular}{p{0.13\linewidth}|l|p{0.26\linewidth}|p{0.2\linewidth}|p{0.1\linewidth}|p{0.09\linewidth}}
        \textbf{Tool} & \textbf{Creator}  &\textbf{ Motivation} & \textbf{Alt. Approach} & \textbf{Mode} & \textbf{App/Web} \\ \hline \hline
        Car identifier & \AZ*, \KL & Support more reliably finding Uber & Calling Aira or a sighted friend & Live & App \\ \hline
        Mail sorter & \AS & Distinguish between relevant and junk mail & Take repeated photos with BeMyAI & Live & Web \\ \hline
        Hand gesture interpreter & \GK & Better understand how hand gestures are used in practice & Call AIRA & Streaming & App \\ \hline
        Playing card reader & \GK & Understand a changing hand of playing cards & Use a tactile set of cards & Live & App \\ \hline
        Braille translator & \AJ & Support non-Braille readers in interpreting Braille content & N/A, he personally already reads Braille & Streaming & App \\ \hline
        Sauce and seasoning finder & \AZ & Properly identify and locate sauces, seasonings in his kitchen & Repeated BeMyAI descriptions & Streaming & App \\ \hline
        Business card reader & \AS*, \AZ & Read only the relevant text on a business card & SeeingAI & Streaming & Web \\ \hline
        Chess move validator & \KL*, \GK & Validate moves made in a game of chess & Streaming board to sighted friend & Live & App \\ \hline
        Vacuum assistance & \GK & Determine if a spot is clean or has obstacles in it while vacuuming & Touch spot with hand and feel it & Take photo & App \\ \hline
        Light detector & \AJ & Communicate the level of light in a room verbally and via earcons & For him, SeeingAI. For a deafblind person, no solution & Streaming & App \\ \hline
        Bus stop detector & \GK & Help identify if there is a bus stop and where & Use maps to get 95\% of the way, from there, ask someone & Streaming & App \\ \hline
        Scale reader & \AZ & Read the weight on a kitchen scale in grams & Talking kitchen scale & Streaming & App \\ \hline
        Room name identifier & \AZ & Read aloud the names on room signs & Run hand along wall to read braille room signs & Streaming & App \\ \hline
        Straight line walker & \AS & Help guide to walk in a straight line & N/A, veers when walking & Streaming & Web \\\hline
        Makeup checker & \AS & Check makeup is applied correctly & Use BeMyAI & Take Photo & Web \\ \hline
        Clockface food description & \AJ & Describe where food is on a plate using clockface descriptions & Poke around with fork & Live & App \\\hline
        Waffle plushy finder & \AZ & Find plush dog (Waffle) if in frame & N/A, exploratory not assistive & Streaming & App
    \end{tabular}
    \vspace{1pc}
    \caption{A sample of tools created in our study. Table includes a description of the tool created, the co-designers' alternative or current approach to accomplish the same task (i.e., what strategy they used without the bespoke AT, if any), the mode of the tool (live, streaming, or photo), and how the tool was created (via the app or web interface). If multiple co-designers made the same tool, details correspond to the co-designer marked with asterisk. Tools built but not listed here can be found in \Cref{tab:tool_tableappendix}}
    \label{tab:tool_table}
\end{table*}

\subsubsection{What tools did co-designers create?}
Our five co-designers built 37 tools in total (see \Cref{tab:tool_table}) to receive visual assistance with navigation , retrieve information from physical objects, and for social factors including self-presentation and to understand physical social cues. We asked our co-designers for a specific number of tools (at least four per phase), but did not constrain what they built beyond being camera based. Co-designers had many ideas they were eager to try, often exceeding what time allowed for: making a total of 37 tools, exceeding the minimum expectation of 28 tools. We present their motivations for building these tools, and discuss their own considerations to assess tools' success. 

Some tools correspond to common, well documented use cases where the motivation was straightforward: they assisted with an unavoidable task, or there is either no pre-existing, efficient, commercially available AT for the task. For example, tools for outdoor navigation included tools for solving well-studied accessibility problems of finding bus-stops and room numbers, and for specific needs to follow a straight path when walking. Tools for information access followed similar motivations: two co-designers built tools to extract information from business cards—a ubiquitous part of networking. For these types of tasks, even while building bespoke tools, co-designers often aimed to solve similar problems. All five co-designers expressed an interest to build tools to find an Uber, and two co-designers attempted to build this.

Other cases were more niche: examples include tasks already addressed by mainstream AT (e.g., \AJ's light detector, which has an analogous SeeingAI mode), tools related to longer-term social understanding over immediate use (e.g. \GK's hand gesture interpreter) and tools for tasks that are nominally already accessible (e.g. \AJ's Braille reader). These tools often share the motivation of lacking a pre-existing AT to address it, but are more diverse in their alternate motivations. \AJ's tool, for example, were often motivated by needs outside his own: while SeeingAI already has a light detector, it uses tones only, which is adequate for him but unusable for someone who is deafblind, prompting \AJ to make one that offered spoken descriptions as well because those could be rendered by a Braille display. Similarly, though \AJ reads Braille fluently, he was motivated to build a Braille reader to support Braille learners, or even sighted people navigating the estate of a blind loved one.

\subsubsection{How do tools of interest change over time?}
Co-designers’ creation trajectories began with highly pragmatic, need-driven tools, but diversified over time as they became more comfortable with \studyinstrument and its capabilities. Early tools tended to focus on immediate, functional challenges reflecting a prioritization of utilitarian value. Examples of first tools include a tool to locate cooking ingredients made by \AZ and a tool to locate commonplace physical items made by \KL. However, throughout the study's duration, the tools co-designers were drawn to became more varied, expanding into contexts of entertainment—as found in \KL's final tool, a chess move validator—
or even proof-of-concepts for AI workflows that were not themselves AT. \AZ's second-to-last tool tested a pipeline for generating an image description that a \studyinstrument tool could then use as a personal object recognizer. 
\GK explains this evolution arises from exhausting some more basic use cases, saying: \Quote{you know, most of those basic things, like, the more fundamental things, I feel like I've gotten a good grip on in my life. So, like, what are other, like, areas where I feel like I might not necessarily need AT, but my life could potentially be enhanced by AT, which ended up being a lot of \ldots info that I didn't necessarily `need-need', but could benefit from.} 
As time went on, and the floor of needs could be addressed, co-designers felt comfortable to explore niche, nice-to-have use cases for which they may not have even considered AT in the past.

\subsection{How Are Tools Built?}
\textbf{There was no single path to successfully building bespoke assistive tools: co-designers found success through different prompting and testing strategies, including both carefully engineered one-shot prompts and progressive iteration. Across these approaches, failures were largely explainable, arising from technical limitations, AI capabilities, or misalignment between specifications and implementation.} 
\subsubsection{Characterizing prompt strategies}
All five co-designers indicated that their strategies for approaching the initial creation of tools, as well as their iterations influenced the tool's ultimate success. Yet, all indicated different strategies that corresponded to ``best'' results, according to each co-designer’s definition of success. \AS found the most success when treating her prompts for tool creation as prompts to the code agent directly rather than requests for a given tool, generally being explicit about technical details and desired input and output, and adopting lengthy, highly structured initial prompts: consistent with standard prompt engineering practices in software engineering environments \cite{dicuffa2025exploring}. Through this practice, \AS had multiple tools achieve success with one iteration or fewer: an accomplishment not paralleled by any other strategy she tried. \KL, however, found more success through abandoning this approach. Though he started with long and detailed prompts, and even manually selecting the agent's model, lengthy prompt-crafting and code generation times urged him to try something simpler, and he ultimately found shorter, and action-oriented prompts to be more successful: \Quote{Well, [initially] I did, like, a long prompt, and then also I used, I changed the model from Auto to GPT, and it took forever to [generate the code]\ldots also, even though it took such a long time, it didn't work too well, and and so, I changed it to auto at the end, and then just gave it a very short prompt on what I actually want, and then it kind of worked after that. So my initial assumption is that long, detailed prompts with a good model will work, but \ldots that wasn't the case \ldots, that short comment, it just started working way better.} He notes that if this abbreviated approach lacked any specific detail, \Quote{I can keep on reiterating with short comments, rather than making a long prompt with everything I want, and then all the edge cases I could think of.} This approach aligns with the practice of progressive specification \cite{tang2026programming}. 

The preferred approaches of \GK and \AZ are similar to that of \KL in their brevity and avoidance of hyper-specification, but motivated differently. For \GK, successful tool prompts often started broad, in an interest to see the possibility space of what the models underlying \studyinstrument could provide, and using iteration to optimize for where the model was strong. In creating his most successful tool, the hand gesture describer, he described his process as \Quote{initially\ldots somewhat broad,\ldots like here's the motivation, here's the challenge,  help me solve it. And then, after testing it, \ldots I got a sense of what the descriptions might look like\ldots and the rest of the iterations were latching on to that detail.} When tailoring his tools to leverage what the models could accomplish rather than fighting a model's weakness, \GK found tools exhibited less friction. 

For \AZ, by contrast, avoiding hyperspecification was motivated by being selective on what the model prioritized. His prompts early in the study were typically done via dictation in the mobile interface, producing a stream-of-consciousness specification containing all thoughts pertaining to the tool without an implicit sense of prioritization, contributing to a very low success rate for phase I tools (0/5). For his most successful tool—the car finder—, however, \AZ instead used a text file to author a detailed, but intentional prompt, which he then pasted to the mobile interface. He described his rationale for doing so as \Quote{I don't want to follow the template, that's just too many steps\ldots, but I also don't want to dictate, because this has very specific things that I want to hit, and \ldots if I'm just talking a lot to the AI in the form of a dictation I might include something that, to me, is just me running my mouth but to [the agent] is a Bible verse.} Though \AZ found the web interface's templated structure tedious, favoring the expressiveness of pure natural language, being judicious with included information yielded a tool that met his expectations. This result is consistent with literature indicating that, while appropriate context is vital to successful AI interactions, providing too much detail dilutes model focus \cite{avoid-overspecification}.

For \AJ, the favored strategy was explicitly tied less to what to include than how to include it: creation often happened quickly via natural language whereas iteration was a more deliberate process that invited closer scrutiny and sustained attention. In effect, prompting from mobile, but iterating on the website to take advantage of the real time logs became his preferred workflow; \AJ describes \Quote{I love the mobile because\ldots I was able to get my thoughts down \ldots fresh. But, like, when I make iterations to it \ldots I will do them from the GitHub website, so that I can actually monitor what it's doing, and if I see something that it's doing that I'm like, maybe it should do something else. I could just send a follow-up prompt.} Materially, this aligns with the tendency towards progressive specification \cite{tang2026programming} highlighted by \KL and \GK, but also highlights the relevance of input modality in surfacing these strategies.

Across creation strategies, co-designers placed strong emphasis on testing and iteration as core parts of their tool-building process, repeatedly trying tools in real-world contexts and using those experiences to guide refinements. Rather than treating evaluation as a separate phase, co-designers folded it into an ongoing loop—testing, reflecting, and updating prompts to better align system behavior with their needs. \GK refers to this strategy as ``experience reports''—a process by which \Quote{I try to give it specific \ldots reports \ldots of what went wrong to try and correct the functionality\ldots just, like, how it failed, basically, `hey, like, I tried this, this is what the output was, it's not what I wanted. I was expecting this, go fix it.}. Experience reports were a common practice amongst co-designers: efficiently surfacing breakdowns which informed concrete changes in subsequent tool versions. In doing so, participants effectively enacted a full human-centered design cycle, where evaluation and iteration were continuous and user-driven rather than discrete steps \cite{ladner2015design}. 

Testing practices were also frequently shaped by the context of the tool itself. In particular, for tools that processed information considered private by co-designers, privacy often informed testing practices. One example of this is \AS's mail sorter tool (described in \Cref{tab:tool_table}. Since mail, junk or otherwise, generally contains sensitive information such as one's home address, \AS was very particular about what criteria should be used in analysis, and what information should be read to perform such analysis. Her tool prompt made this criteria and scope explicit. While her hope for the tool was eventually that it would help her bulk process mail she had not seen, she first tested it with a few known pieces of mail, to see how it would respond and verify it honored by her criteria. Testing first with known examples was a frequent technique across many tool contexts, but \AS felt it specifically important here because it offered observability; she notes \Quote{I am a security person, so, like, what are the security privacy implications of this\ldots if I'm kind of running something that I don't fully understand, I'm running an AI authored tool, and it is also sending my picture to some other website behind the scenes? \ldots That's where, like, adding more transparency and observability controls really helps, right? Like, anything AI that you build, it's good to have, like, more observability controls so that your user can see what exactly happens, and can actually build trust with the tool.} Though testing with known examples is not always possible, nor is it a foolproof answer to questions of observability, it served as a valuable step towards building trust with tools, especially those with clear potential for privacy risks.

\subsubsection{Examining contributors to tool success}
Tools that stood out most positively in our study did not simply succeed because they ``worked.'' Instead, they exceeded a baseline expectation of functionality by shifting what was possible or practical in participants' everyday workflows—either by introducing new capabilities or by dramatically reducing friction in tasks they already perform.

Some standout tools addressed needs that had no accessible or reliable implementation through existing consumer technologies. Participants described these less as incremental improvements and more as qualitatively new forms of access. \GK noted that, while his hand gesture tool was one of several successes, \Quote{what made it different for me was if this tool didn't work then\ldots I could ask sighted people or call AIRA, but realistically, I'm not gonna do that.} \AZ drew a similar distinction between his scale reader—a successful but replaceable tool—and the Uber car identifier: \Quote{I could just get a talking kitchen scale for 20, 30 bucks, whatever. I can't get a \$20, \$30 car identifier off of Amazon. And that's why this tool stands out, because it is so unique.} In cases like these, even imperfect performance was often tolerated, as the alternative was having no comparable option.

Other tools derived their value not from novelty but from reducing cumulative friction in the tasks co-designers already performed frequently. \AS, for instance, could sort mail using BeMyAI, but doing so in bulk was cumbersome. A custom tool transformed the experience: \Quote{I think [mail sorting is] one of those things which I generally do in bulk\ldots I'm sitting with a whole load of things, and just trying to sort through that, so I feel like success feels even more profound.} Novelty, in other words, is not a prerequisite for impact.

Across both categories, definitions of success were pragmatic rather than absolute. Tools were considered successful when they were reliably good enough to support a real-world goal, even if imperfect. Reaching that threshold, however, rarely came from a single ``correct'' prompt. Instead, it emerged through iteration: adjusting specificity, structure, and constraints in response to observed failures. \GK rescoped a general video description tool into one focused on hand gestures, finding a niche where it fit. \AS dug into the codebase to discover that her rationale for what made mail worth opening had never been included in the repeated query because it was omitted from the \texttt{custom\_gpt} template field.

For co-designers initially resistant to iteration, eventual engagement proved transformative. \AZ describes his scale reader as a \Quote{turning point\ldots at which I started to approach toolmaking differently, where I'm like, okay don't expect perfection from one shot. You need to meet it in the middle.} The tools he built afterward—the Waffle plushy finder and the car identifier—were rated among his most successful. Working tools, then, reflected not just a final prompt but the process through which it was shaped: experimentation, observation, and incremental refinement.

\subsubsection{Examining tool failures}
Tool failure generally comes from one of three places: requiring a technical approach \studyinstrument does not support, failure of a model a tool utilized, and misalignment between specifications for a tool's behavior—provided by both the co-designer and by \studyinstrument's structure—and how a tool ultimately performs. 

The first case is simple to identify: tools that hinged on a functionality not yet addressed by \studyinstrument tended to fail. For instance, \AS's straight line guide relied on the ability to cache a starting photo and use it as a comparison for all future image inputs; without explicit support for persistent memory and a cache, this was unlikely to succeed. Co-designers avoided this pitfall through two primary techniques. The first is referencing the repository's README and architecture files to get a sense for how \studyinstrument works under the hood and adjusting their requests accordingly. \KL took this approach in iterating on his stain finder tool, describing that he \Quote{changed the prompt based on the README and architecture description. So it made it more specific, because I think the prompts before I wrote were saying some, very unnecessary things that might not be actually possible with this}. The second technique is creating tools in a way that circumvents a technical limitation entirely. For example, since uploading one's own data is not yet supported, \AZ asked ChatGPT to describe a specific object—a plush dog named Waffle—and used that description in a prompt to \studyinstrument, building a personal object recognizer without the need for local training that would typically underpin such a tool.

The second case, failure of a model that underpins a tool, proved harder to pinpoint as a source of blame for a tool's failure since there is no visible line between where the code stops and the models it calls begin in the mobile interface. Co-designers who identified this as a source of tool failure typically did so by trying to accomplish a similar task through other, general-purpose AI platforms like ChatGPT and seeing if the problem persists: if it did, they attribute the source of failure to AI, and if it did not, they attribute failure to some characteristic of the tool itself. For instance, in his door finder, \AZ found that even when the tool built with \studyinstrument successfully identified whether a door's hinges were visible, it could not reliably map that to the direction the door opens. \AZ found this debugging strategy useful because \Quote{it's easier for me to\ldots understand the boundaries before I can understand the capabilities\ldots left-right is already a difficulty of AI so I'm not going to fault it.} \AJ had a similar experience while debugging his Braille translator. After the tool hallucinated a Braille translation, he tried ChatGPT and found that it would not even make an attempt, saying \Quote{It refuses. Straight up refuses. ChatGPT will not mess with Braille. It'll give me braille symbols\ldots but that's all it will do.} Though different from \AZ's experience, this also serves as evidence that tools' failure is often a result from asking AI to do a task it is not trained well or at all on.

The final case, misalignment between specifications for tool construction and behavior and the tool's actual implementation and performance, was often the most subtle: arising not from clear technical limitations or model failures, but from gaps between how co-designers intended a tool to behave and how those intentions were interpreted by \studyinstrument. Our code-review analysis reveals that clearly missing or misinterpreted user specifications were somewhat infrequent: the code's documentation after initial prompting reflected user-provided specifications in all but five tools. Instead, violations of user specifications tended to occur due to \emph{spec conflict}: where two specs for a tool cannot mutually co-exist. One example of this was found in \AJ's food finder tool, which aimed to use clock-face directions to describe where on a plate different types of food were located: while the documentation reflected clock-face directions, Copilot's instruction file in the \studyinstrument repository bounds clock-face directions to only values on the upper half of the clock, because in the classic navigation use case, values on the bottom half of the clock are nonsensical output because they are behind the user's camera. This conflict between user specs and repository level specifications worked to the tools' detriment. 

Spec conflict also manifests in iterations that provide new criteria for a tool with implementation that violates existing specifications. For example, when \KL made his car finder tool, he was happy with its overall output and accuracy but found it slow, prompting him to iterate to improve the tool's speed while holding the rest constant. Copilot latched on to the speed request, refactoring the tool to utilize lower-latency, lower-generality models like YOLO over the existing model in use, Gemini. He described the outcome, saying \Quote{After that comment, it got really fast at announcing whether a car is visible but the accuracy got really bad}, conceding that while Copilot did comply with the new specification provided, it contributed to the tool's failure by doing so without consideration to the other specifications at play.

\subsection{Perceived Value of \studyinstrument}
\textbf{Overall, co-designers viewed \studyinstrument as more than a development environment: it served as a testbed for experimentation, a catalyst for re-imagining accessibility, and an accessible entry point to creating custom technology.}

While co-designers found value in the tools \studyinstrument facilitated, this was not the only benefit of \studyinstrument. Beyond a development environment, co-designers thought of \studyinstrument as a ``testbed'', a ``thought exercise'', and a ``bridge'' to creating for oneself. 

As a testbed, co-designers framed \studyinstrument not just as a way to build custom tools, but as a means of testing what is possible. \AZ frames this as \Quote{a sanity checker.\ldots Like, okay I have this idea, lets put it into practice immediately, a reality check of sorts, what’s possible and what’s not}. This perspective persists even when what's possible does not translate to a fully idealized tool; he notes \Quote{you still have to start somewhere, and \studyinstrument is the start.} Even for co-designers whose sense of value comes primarily from the utility of created tools, the availability of seamless, rapid testing was prized. As \AS puts it, \Quote{testing was the most delightful aspect, how easily I was able to just load a PR} The ease of initial creation, and low time investment in creation serves as a low-risk way to test interesting and challenging ideas; as \KL notes \Quote{I mean, it's just, so, so easy, right? Opening up a computer and, writing an actual program, compared to just, like, pull out the phone and typ[ing] a few sentences}, which he found de-risked ambitious creation.

In contrast, the framing of \studyinstrument as thought exercise is reflective—rooted in the ability to positively disrupt habit. \GK notes that, prior to becoming involved in the project, he was not always thinking about places AT would fit, because he already had workarounds in place: he notes that \Quote{a lot of blind people, including myself, end up getting very used to the way that we live currently, especially if we've, you know, figured out non-visual workflows for things. So it's\ldots hard to kind of be like, oh, imagine a world where you could make improvements to your life or to your workflow.} The study offered an opportunity for \GK to \Quote{exercise [his] imagination} and explore \Quote{what if, like, there was a different approach to this, or what areas in my life could I maybe try to make things even more efficient?} Notably, this application had value even when technology was not ultimately the answer, as occurred while making a vacuum assistance tool, which he \Quote{ended up not needing to use\ldots, doing this my own tactile way was more efficient than aiming the camera, waiting for the description, etc.} There was value in the exploration and probing the what-if, as he puts it \Quote{the more valuable part is to start getting out of the \ldots I don't want to call it subpar, but I'm so used to the techniques that I'm using right now that it's hard to, like brainstorm or imagine.}

For others, the added value was not just in the freedom to imagine, but the freedom to do. While all co-designers were recruited for their technical experience, not all are formally trained programmers, leading some, like \AJ ,to view \studyinstrument as a bridge to creating for oneself. \AJ, not a programmer by trade, noted that \studyinstrument offered support for the entry into customizing code to meet their own access needs, in particular because the underlying structure was already established, saying \Quote{especially for somebody who is new to coding, if they were working with this app, getting experience in developing their own kind of setup with a mainframe that's already there.}

\subsection{Supports and Structures to Create Bespoke Visual AT}
\textbf{Overall, while co-designers saw conversational programming as lowering barriers to creating bespoke assistive technology, they emphasized that democratization requires more than natural language alone. They identified conversational guidance, collaborative tool sharing, and support for user-provided data as critical structures for enabling broader participation and higher-quality tools.}

Co-designers unilaterally saw \studyinstrument as something that democratizes creation. \AZ expressed a desire to show \studyinstrument to his non-technical friends, believing it would empower them to create for themselves: \Quote{I have some average Jill and Joe friends. Blind people that we do our research for, not academics. I would like to show them \studyinstrument, and sit them down, have them give a one paragraph description, and see what they make\ldots I think even that audience can\ldots really make their own proof of concept. Most of these people don’t spend much time on a computer, they use their phones\ldots, so that experience alone makes programming accessible to all BLV people, whatever they do for work.} \AS echoed this sentiment, expressing that \Quote{the more democratized we can make it, the more people can actually leverage... I feel what \studyinstrument does well is democratize the experience of building assistive technology\ldots I'm pretty sure it would feel amazing if I wasn't a developer to get access to something like this, because, oh my god, I suddenly become a creator, which most people can't.} \GK extends this line of thought, adding that systems like \studyinstrument could support less technical people in feeling comfortable with technology; he imagines \Quote{people might be able to get a sense of, like, I'm able to leverage this, and I'm not as scared about AI and technology as maybe I used to be.} While future work is necessary to validate the utility of conversational programming for bespoke camera based AT in less technically savvy BLV populations that our co-designers predict, these perspectives nonetheless point to its potential as a pathway toward broader participation in assistive technology creation and a shift in how users engage with and feel empowered by AI-driven tools. We present findings that discuss appropriate supports to engage technical and non-technical users alike in building their own bespoke AT.

\subsubsection{Creation as a conversation}
Multiple co-designers noted that expanding the notion of conversation to the creation step could empower more intentional tool creation and be a valuable scaffold for those who may not know precisely what they want. Currently, conversation begins mostly at the iteration step, implicitly expecting the user to know precisely what they want; \AZ describes, \Quote{You kind of have to come with a full idea. It makes us more of the thinker, and it's just the tool.} The ability to sit, think this through, and arrive at a conclusion is well suited to more formal prompting, as the desktop experience enables: \AZ notes that \Quote{[he does] really enjoy sitting than writing it down and having time to think \ldots and actually come up with things that are worth doing}. Yet, especially for mobile prompting, where the experience is more on-the-go and flexible, some of this structure for thinking through precisely what is desired can be lost. To this, \GK—who prompted almost exclusively from the mobile interface throughout the study—recollects that \Quote{especially with the initial stages, for me, it was like, well, I have this idea, and then I'm not entirely sure, like, how the AI system is going to, like, respond...it could have been very nice to, before AI goes and runs off and does its thing, check in with me and kind of set up this mutual understanding [and] give me a chance to riff off of whatever the AI is, thinking about \ldots before I actually test out the tool.} Bringing a conversational flow to the creation step could support end-users in both refining what they want out of a tool and optimizing their request to simultaneously meet their needs and align with what the underlying AI is capable of. However, adjusting this structure would come with tradeoffs. Simplicity of expression is one of the most appreciated aspects of the \studyinstrument creation experience, and introducing a conversational layer could add friction. For \KL, \Quote{even though it's just answering it back, it's still not as simple\ldots I'm not sure how people would feel about that.}

\subsubsection{Bespoke, yet collaborative}
\label{subsubsec:collab}
While all co-designers found being able to create unique tools valuable, some  also mentioned a desire to be able to share the tools they made with others around them. While there are many needs not covered by commercial AT, the challenges they experience are typically not so unique others would not benefit from the solutions they build. This intuition was validated by the tools co-designers made over the course of the study: over 20\% (8/37) of tools created were shared across participants.
\GK expressed that while he could try to support someone he knew in building a tool for themselves by describing the steps he took to do it, \Quote{you're still gonna have to do those steps for yourself, and maybe that's a good thing, so they can personalize it, maybe my tool is not exactly what they're looking for, but at the same time, I feel like there's still this issue of people duplicating efforts, and, like, because of maintenance and scalability and just, like, sustainability.} \AS echoed this sentiment and advocates for a path to share the progress she has made with others, saying \Quote{leveraging the power of community can be a very ideal next step, like how do we build on each other? Like today I built the mail sorter tool, let 10 people use it and iterate on it and then we all benefit. We don’t just build for ourselves.} \AS, as the co-designer who most frequently used the GitHub web interface, would occasionally see activity in other co-designers' repositories, and found that this could offer a source of inspiration for things she would find useful herself, further grounding her collaboration would add value to a platform  like \studyinstrument, noting that \Quote{as I was seeing other things that others have created, then I thought, `hey, that can be useful to me'.}

\subsubsection{From do-it-yourself tools to bring-it-yourself data} 
A recurring theme across co-designers was the desire to bring their own datasets into the tool creation process, particularly for tasks that require a high degree of specificity. While participants generally felt that existing tools performed well on broad, general-purpose tasks, they noted clear limitations when attempting to address more niche or context-dependent needs. In these cases, relying solely on non-specialized language models often resulted in outputs that were too generic to be useful, thus, co-designers expressed interest in augmenting the system with their own data, enabling tools to better reflect the particular environments, objects, or patterns that mattered in their daily lives. \AZ references this as a struggle of tools based on generalized LLMs, and finds navigational tasks to be an area where this is particularly salient, expressing that \Quote{If it's based on an LLM, it's really cool, but I always get at least one hallucination. I much rather prefer if there's something that I'm…using the AI for be trained for that specific purpose\ldots with like OCO: it's simply just trying to tell you whether or not a pedestrian signal is telling you to cross or not.  When it's LLM-based, I find it struggling in trying to do a lot of things at once. Jack of all trades, master of none.}

Beyond simply improving performance, this desire points to a broader distinction between generalist and specialist tool use. Participants recognized that the current infrastructure excels at quickly scaffolding widely applicable functionality, but saw opportunities to extend this capability by incorporating more tailored sources of knowledge. This included not only uploading custom datasets, but also the possibility of integrating or switching between models better suited to specific domains. 
For example, in trying to build a braille reader tool \AJ found general purpose models prone to egregious hallucination, and hoped to use Braille specific models to remedy this. Likewise, \KL hoped to provide a dataset of photos of chess moves to support a chess move tracker that could better understand what constituted a valid move. Co-designers attempted workarounds to this limitation, such as attaching photos to the GitHub issue for the tool to use as reference, as demonstrated in \AZ's car finder, which made some progress, but served more as a primer to imagining what the bring-your-own-data experience could look like than a functional stand in. This was expressed by \AZ, who notes \Quote{It would be nice if I could just, like, sideload a model, like, for example, like, color reading, let's put a model, QR code reading or even image description, you can do it in \studyinstrument.\ldots All possible.} Such flexibility was framed as a way to bridge the gap between rapid prototyping and sustained, high-quality use, allowing tools to evolve from general approximations into more precise and reliable supports for specialized tasks and enabling utilization of \studyinstrument as an ecosystem that can address a wide and comprehensive variety of needs, rather than a singular AT artifact.

\subsubsection{AT co-use}
All co-designers, prior to the study, used existing AI-powered AT such as BeMyAI and SeeingAI, and articulated the cases in which they would reach for one type of tool over another. Generally, co-designers agreed that \studyinstrument tools were preferable for specific, repeatable tasks. Likewise, pre-built tools were preferred when generality was desirable, or when the task at hand would benefit from the shared experience of others, since the pre-built tools have thousands of users, while any \studyinstrument tool is personal to the individual. 

While, today, these use cases are largely separate, co-designers also speculated on futures for what co-use of bespoke tools like those made with \studyinstrument and pre-built tools like BeMyAI or SeeingAI might look like. One prominent perspective envisioned a complementary relationship, where general-purpose tools and custom solutions work together to balance breadth and specificity. In this view, pre-built systems would continue to handle a wide range of common tasks, while custom tools would fill in gaps or provide more tailored functionality when needed. Participants described several ways this synergy might manifest. \GK proposed a model where general systems can infer user intent based on visual context and dynamically route tasks to the most appropriate tool, including custom ones.  \AS visualized something similar, but where an interface akin to \studyinstrument was integrated in more generalized tools, allowing for the creation of bespoke tools as plug-ins as a way to address that no general tool can possibly meet every need. She argues \Quote{BeMyEyes or SeeingAI should have, like, these kind of concept of plugins, right?  Like, most AI tools are now letting you define your own tools, or define your own custom agents and things like that. \ldots In accessibility you cannot always check all the boxes. Every person is different \ldots and people know what works for them, so let them do that.} In contrast to this fully integrated approach, \AJ envisioned a future where pre-built and bespoke tools still exist separately, but can be freely surfaced on top of each other using mobile automations, similar to the concept presented in \etal{Herskovitz}'s A11yExtensions \cite{a11yextensions}.

Other co-designers anticipated a more competitive trajectory, where advances in one category eventually render the other less relevant. \AZ remarked that while \studyinstrument is not yet advanced enough to unseat SeeingAI, when \Quote{\studyinstrument gets to the point that it has a sleek, sexy, straightforward interface, similar to those of Seeing AI and Lookout \ldots I will drop seeing AI, because I can make what Seeing AI does in \studyinstrument\ldots like, \studyinstrument is not one assistive technology. It is an AT ecosystem.} Under the ecosystem framing, \AZ believes that with time and technical development, \studyinstrument could supplant existing tools by allowing end-users to reproduce the functionality of pre-built tools in a way that works best for them. While not every co-designer envisioned a future of co-use, some also speculated that mainstream AT could render the need to create AT conversationally irrelevant, or vise-versa. \KL noted that \Quote{If one gets good, then I think the other will probably lose the market share}, which could come either from the refinement of systems for custom tool creation as \AZ describes, or from model improvements resulting in pre-built tools being more capable of addressing niche use cases.

Still, even with a future including these supports and structures in mind, some co-designers called into question whether democratizing the initial act of creation would successfully empower more people to create useful, usable tools. \GK expresses that though the benefits of more people being able to prototype and tailor tools to their needs are clear, the promise of this ultimately yielding more high quality tools is not. He remarks \Quote{I don't know, it's a little bit unsure, still, if, prototyping easier and faster just means that we get more quality final systems\ldots It's easy to quickly iterate on stuff\ldots but I just don't know high the ceiling can be raised. I think people would just kind of tend to use janky-ish technology, like their own personal tools and\ldots once people get into a rhythm of this is how I'm using AI in my life \ldots then you just kind of get stuck there, and you might not be able to have newer ideas with newer technologies.} Even in finding \studyinstrument valuable as a thought exercise, it too runs the risk of forming suboptimal habits rather than challenging them. He notes the risk of forming suboptimal habits and suboptimal tools would likely be more pronounced in the same community it stands to empower most—users without extensive development experience, because \Quote{after you make a tool, and then try to iterate on it, but then if it's not working, then what happens? And I feel like especially for people who aren't…used to thinking about programs, I think that's gonna be more frustrating and more challenging in a lot of ways.} In order to take advantage of a system like \studyinstrument's anticipated power to democratize creation, future work must better understand and support the needs of novice creators.

\section{Discussion}

We examined how BLV co-designers engage in the creation of bespoke, camera-based assistive tools through an agentic programming system, focusing on both the conditions under which tools succeed and the sources of their failure. Our findings highlight that while such systems can enable rapid and flexible tool creation, success depends on a combination of factors, including alignment with system capabilities, the reliability of underlying models, and the co-designer iteratively refining prompts. Participants were broadly optimistic about the potential of \studyinstrument to expand access to personalized assistive technology, while also surfacing key challenges in usability, transparency, and robustness. In this section, we consider these barriers and opportunities, and translate them to recommendations for future platforms, grounded in co-designers’ experiences (\Cref{subsec:rec1}, \Cref{subsec:rec2}, \Cref{subsec:rec3}). Then, we discuss the limitations of these findings (\Cref{subsec:limitations}).

\subsection{Expand Creation and Access Modalities}
\label{subsec:rec1}
\studyinstrument supports multiple modes of creation  and iteration of visual assistive tools, yielding nuances in how these modalities influenced our co-designers' approach and motivations to building their tools. The templated interfaces available via GitHub enabled co-designers like \AS to deliberately compose and version tools outside of an immediate use context—affording time for reflection and structured refinement—whereas the free-form interface on the app supported a more in-situ and unstructured approach to quickly converge a situationally-inspired visual access need into a functional, testable tool. Though powerful, \studyinstrument only supports testing on mobile devices, limiting in-situ inspiration to situations that do not expand to hands-free use. Expanding \studyinstrument's testing capabilities to support smart glasses could inspire a wider range of visual tools for hands-free scenarios such as cooking \cite{huh2025vid2coach, franklin_cooking}, where visual assistance was desired by multiple participants (\KL, \AZ), but since the hands were occupied with the task itself, it was impractical to use mobile assistance.

Further, supporting testing through smart glass form-factors warrants the need for intent specification modalities that are also hands-free. We look to making AT creation more conversational as one potential avenue. While iteration proved effective, the current workflow assumes that users begin with a well-formed idea of what they want to build. Several co-designers noted that this places a cognitive burden on the user, particularly in early stages when goals may still be evolving. Enabling back-and-forth interaction before a tool is generated could help users clarify their intent, explore alternatives, and align their requests with what the system is capable of, rather than committing prematurely to a fixed specification. When working from a hands-free, keyboardless interface like glasses, the challenges to appropriate specification via dictation mentioned by \AZ are magnified, further motivating the need to support users in clearly thinking through what they want.

At the same time, integrating conversation into creation introduces relevant design tradeoffs. In particular, the current experience is valued for its immediacy and low overhead, allowing users to quickly express an idea and see results. Adding a conversational layer could risk increasing friction. Future work should explore how to balance these dynamics, for instance, by supporting lightweight, optional dialogue that scaffolds early ideation without disrupting \studyinstrument's appealing speed and simplicity.

\subsection{Collaborative Testbeds to Support Verification, Knowledge and Tool Sharing}
\label{subsec:rec2}

The popular framing of \studyinstrument as ``testbed'' by our co-designers points towards a broader opportunity to not just probe the capabilities and limits of AI systems, but to scaffold collective knowledge about how, when, and why they succeed and fail. In our study, co-designers already treated tools as informal testbeds: systematically evaluating them on inputs with known answers before deploying them in higher stakes contexts. This practice reflects a critical need among BLV users to understand system behavior, especially given the limited non-visual explainability of many AI systems \cite{king_of_knowledge} and the difficulty of diagnosing inconsistent outputs \cite{alharbi_misfitting}. However, these investigations currently remain individual and ad hoc, placing the burden of learning system behavior on each individual user.

We argue for collective testbeds that externalize and aggregate this process. Rather than each user independently rediscovering the same limitations, such infrastructures could support the sharing not just of tools as discussed in \Cref{subsubsec:collab} but also the strategies, assumptions and empirical insights that emerge through their use. Collaborative testbeds conceptually extend the more conventional forms of sharing wherein users exchange tools or configurations \cite{share_work_a11y} by also capturing the reasoning and experimentation that make shared artifacts effective. Our co-designers developed tacit knowledge about which models were most reliable for particular tasks, how to structure prompts, and how to interpret uncertain outputs; making this knowledge visible and reusable could reduce duplicate efforts and accelerate effective tool creation, especially when the suite of available models in \studyinstrument expands.

This vision parallels model evaluation dashboards in non-accessibility domains, which allow developers to compare model performance across task types \cite{rashid2025swe, golnari2026devbench}. A similar paradigm for accessibility could answer questions about model suitability for different tasks, or under what environmental conditions certain approaches break down. Importantly, these insights would be grounded not in abstract benchmarks, but in situated use cases.

Enabling this form of sharing raises important design challenges. Tools must be represented in ways that make their assumptions, appropriate contexts of use, and limitations legible. Mechanisms for versioning \cite{lifecycle}, attribution, and iterative refinement are also necessary as tools evolve across contributors. At the same time, systems must preserve the flexibility for personalization that motivates bespoke tools in the first place. Future work should explore lightweight models of collaboration that balance supporting reuse and collective intelligence without constraining individual adaptation.

\subsection{Agentic Support for Scalability and Sustainable AT Creation}
\label{subsec:rec3} 

Finally, we argue for strengthening agentic supports as a path towards scalable and sustainable AT creation. A first key step is to expand \studyinstrument's set of reusable building blocks. Current components are motivated by core functionalities modeled in existing AT (e.g., scene description \cite{bemyai}, OCR \cite{microsoft-seeingAI}) or functionalities that prior literature suggests would be valuable (e.g., camera aiming \cite{a11yextensions}). These provide a strong foundation, yet, the tools of interest in our study suggest the need for additional primitives such as teachable object recognizers or parameterizable components. Broadening these building blocks would better support diverse tool creation while reducing the effort required of end-users.

Yet, for these tools to scale in utility, we must also consider how \studyinstrument might scale to the broader BLV population, necessitating explicitly addressing the cost structures underlying these systems. \studyinstrument currently relies on a host of paid architectures (e.g., server hosting, API calls, Copilot subscriptions) which limits scalability, particularly within disability communities that disproportionately face economic barriers \cite{dandona2001socioeconomic}. Supporting open-source development, self hosting, and lower-connectivity use cases are important steps towards equitable access.

We recommend the adoption of small language models (SLMs) to address equitability and sustainability challenges as well as extend agentic support. Deploying SLMs on-device or in lightweight local environments can significantly reduce dependence on paid APIs and robust server architectures \cite{samal2025small}, while simultaneously introducing new opportunities for tools that require lower latency than LLM-enabled tools, but still need more generality than scoped models like YOLO offer \cite{zhou2023mini}. This shift positions SLMs as both a legitimate technical alternative and a key enabler of scalable, sustainable, and community-accessible AT ecosystems.


\subsection{Limitations}
\label{subsec:limitations}
\subsubsection{Sample size}
\label{subsubsec:samplesize}
While our co-designers were intensively involved in the research process, generating rich insights and data, a sample of five remains insufficient to safely generalize to the BLV community at large. The technical experience of our five co-designers may not generalize to all BLV users; non-technical users would likely require supports beyond those surfaced here to achieve similar results \cite{ma2026not}. However, rather than claiming generalizability, we offer rich insights into the practices of technically experienced BLV co-designers, surfacing a grounded understanding of emerging practices, breakdowns, and support needs that can inform future system design and broader investigations.

Still, while five participants is not large for an empirical study, we maintain it is not remarkably small for co-design. Methodologically similar works draw on three \cite{cossovitch, dance_codesign}, two \cite{a11yextensions, gamage_codesign, charta11y}, or even one \cite{socialsensemaking, bespoke_reflections, jeremy_designforone} co-designer. These works often propose design guidelines: common practice in our community.

\subsubsection{Hallucinations and AI reliability}
\studyinstrument relies on generative AI for tool creation and to support several tools' intelligence. As such, there is an inherent risk of hallucinations. Literature suggests that trust and accuracy are especially vital for systems used by BLV users: making these challenges especially critical to address \cite{king_of_knowledge}. Neither code review nor interviews with co-designers revealed hallucinations in \studyinstrument more pronounced than those present in existing forms of AI-AT such as BeMyAI. Still, this does not negate the inherent risk, which we aim to minimize by supporting users in building validation into their tools. Co-designers approached risk mitigation multiple ways: from building confidence-reporting into tools (\GK, \AZ) to designing control flows in which the model would not answer visual questions prior to validating environmental factors like lighting (\KL). Through these affordances, we hope to engender agency in building tools users can trust, while still recognizing the potential for inadvertent harm.

\subsubsection{Design Considerations for \studyinstrument Study Instrument}
Though \studyinstrument proved useful in supporting development of a variety of assistive tools, its capabilities are limited. 

Firstly, \studyinstrument makes the privacy-minded choice to discard images passed by the user once they have been processed. However, this means \studyinstrument supports only stateless tools: ruling out functionalities such as personal object recognizers (though \AZ achieved a similar result through providing an object description in the tool prompt in his Waffle Plushy Finder (see \Cref{tab:tool_table}), and hindering the success of tools that aimed to surface change over time, like \AS's Straight Line Walker (see \Cref{tab:tool_table}).

Furthermore, while \studyinstrument is distinct from commercial AI tools in many ways—most notably in its generation of real code rather than just wrapping LLM calls and in its ability to iterate and refine existing tools—it still has many things in common with existing AI tools. Indeed, several features of \studyinstrument were directly inspired by commercial offerings. The conversation component of \emph{Take Photo} mode was modeled after the BeMyAI follow up question pattern \cite{bemyai}. Futher, \emph{Live} mode was directly inspired by OpenAI's CustomGPT \cite{customgpt}: being introduced to the study instrument after \AZ experimented with it as an alternative way of making custom tools, but found it cumbersome to constantly need to reprompt.

\section{Conclusion}
We examine how technically experienced BLV users engage with agentic programming to create camera based assistive tools, focusing on their motivations and strategies to create and iterate on tools. Our two-month long study, facilitated by \studyinstrument, reveals that generative AI delivers on the promise to meaningfully lower barriers to prototyping and rapid iteration. Successful tool creation, however, remains an inherently collaborative process: reliant on users correcting, shaping, and extending AI behavior. These results highlight both the promise and current limitations of AI-supported AT creation—despite reducing the barriers to initial creation, AI does not eliminate the need for technical judgment and iterative engagement. By exploring where users find value and friction in authoring bespoke assistive tools with agentic code, we provide design considerations for how future systems could support BLV user-led creation in the future, including supporting collaborative dynamics, emphasizing transparency, controllability, and mechanisms for grounding tool behavior in users’ lived contexts.

\begin{acks}
First and foremost, we thank our co-designers and co-authors Gene, Aziz, Ather, Kun and Aditi for their time and commitment to this work. This research would not have been possible without them.
We also thank our reviewers for their time and feedback. 
This material is based upon work supported by the National Science Foundation Graduate Research Fellowship under Grant No. DGE 2241144, the National Science Foundation CAREER Award No. 2442243, and the National Science Foundation HCC Award No. IIS-2516629. Any opinions, findings, or recommendations expressed here are those of the authors and do not necessarily reflect the views of the National Science Foundation.
This work was also supported in part by a Google Research Scholar Award. 
\end{acks}

\bibliographystyle{ACM-Reference-Format}
\bibliography{sample-base}

\appendix
\section{Copilot instructions}
\label{appendix:instructions}
\colorlet{punct}{red!60!black}
\definecolor{background}{HTML}{f6f6f6}
\definecolor{delim}{RGB}{20,105,176}
\colorlet{numb}{magenta!60!black}

\lstset{
  basicstyle=\scriptsize\ttfamily,
  keepspaces=true,
  basewidth=0.5em,
  frame=single,
  breaklines=true,
  numbers=none,   
  breakindent=2em,      
  breakautoindent=true  
}

\lstdefinelanguage{json}{
    basicstyle=\normalfont\ttfamily,
    numberstyle=\scriptsize,
    showstringspaces=false,
    breaklines=true,
    frame=lines,
    backgroundcolor=\color{background},
    literate=
     *{0}{{{\color{numb}0}}}{1}
      {1}{{{\color{numb}1}}}{1}
      {2}{{{\color{numb}2}}}{1}
      {3}{{{\color{numb}3}}}{1}
      {4}{{{\color{numb}4}}}{1}
      {5}{{{\color{numb}5}}}{1}
      {6}{{{\color{numb}6}}}{1}
      {7}{{{\color{numb}7}}}{1}
      {8}{{{\color{numb}8}}}{1}
      {9}{{{\color{numb}9}}}{1}
      {:}{{{\color{punct}{:}}}}{1}
      {,}{{{\color{punct}{,}}}}{1}
      {\{}{{{\color{delim}{\{}}}}{1}
      {\}}{{{\color{delim}{\}}}}}{1}
      {[}{{{\color{delim}{[}}}}{1}
      {]}{{{\color{delim}{]}}}}{1},
}

\begin{itemize}

\item From the GitHub agent, tools should be written inside the tools subdirectory which is on the same level as backend and ProgramATApp folders. This is where the code will be drawn from. From VSCode, this does not apply.

\item Tool code generated by the GitHub agent in the tools folder should be written in python, not in Typescript. It should run on server, and have results delivered to device. It should return string or dict (not None, not binary). 

\item From the GitHub agent, write all code in a building block pattern where future generated code can leverage and augment it. Also, use any available building blocks in the generation of the requested code rather than writing everything from scratch. This may involve library functions, which is fine, but each tool should be runnable in its own right as well: it must have a main(), run(), or process\_image() function, which take (image, input\_data) as parameters. Do not assign these names to any helper functions. Only use them if that is the intended entry point.

\item When using building blocks, use their content but do not directly import them as modules because this will not work in the exec context. Build them off each other but have them ultimately be standalone.

\item From the GitHub agent, tools must NOT connect to the backend server or use WebSockets. Tools execute ON the backend server and receive data as function parameters. The mobile app handles all network communication. Tools should only process the image and input\_data parameters passed to their main() function and return results as strings or dictionaries.

From the GitHub agent, ALL tools MUST return audio-friendly output. Tool results are automatically spoken aloud via text-to-speech on the mobile device unless another form of audio is specified. Return values should be:
\begin{itemize}
    \item Natural language strings that sound good when spoken (not JSON, not code, not cryptic abbreviations)
    \item Concise - Long outputs will take too long to speak
    \item Descriptive - User cannot see the screen, audio is the primary interface
    \item Action-oriented - Tell the user what was found/detected/processed
\end{itemize}

\item From the GitHub agent, tools can return simple strings OR dictionaries with 'audio' and 'text' keys for advanced control.

\end{itemize}

\small
\begin{lstlisting}[language=json, caption=Tool structure sample prompt portion.]
# SIMPLE (recommended for most tools):
def main(image, input_data):
    # Process image...
    return "I found 3 red stop signs ahead"  # Spoken via TTS

# ADVANCED (for custom audio behavior):
def main(image, input_data):
    # Process image...
    return {
        'audio': {
            'type': 'speech',       # 'speech', 'beep_high', 'beep_low', 'success', 'error', 'warning'
            'text': 'Warning! Obstacle 5 feet ahead!',  # What to say
            'rate': 1.5,           # Speech speed multiplier (optional)
            'interrupt': True      # Stop current audio (optional)
        },
        'text': 'Object detected at 5ft - WARNING'  # Visual display text
    }

# EXAMPLES of good audio-friendly returns:
return "No objects detected"
return "Found 5 people in the frame"
return "Text says: Welcome to the building"
return "Warning: Low light conditions"
return "Scanning complete, 12 items found"

# EXAMPLES of bad returns (not audio-friendly):
return {"count": 5, "type": "person"}  # Don't return raw JSON
return "obj_det_conf_0.95"  # Don't use abbreviations
return None  # Don't return None
return ""  # Don't return empty strings
\end{lstlisting}
\normalsize

\begin{itemize}

\item From the GitHub agent, when tools detect critical information (obstacles, warnings, errors), use the advanced dictionary format with appropriate audio types:
\begin{itemize}
    \item Urgent warnings: `{'audio': {'type': 'beep\_high', 'text': '...', 'rate': 1.5}}`
    \item Errors: `{'audio': {'type': 'error', 'text': '...'}}`
    \item Success confirmations: `{'audio': {'type': 'success', 'text': '...'}}`
    \item Normal information: Just return a string (auto TTS)
\end{itemize}

\item When writing code from the GitHub agent, avoid GPU based packages unless strictly necessary. 

\item For certain types of tasks, there are preferred models unless the user has specified otherwise. 
For object detection generally, this is to use Yolo11 and COCO. For detecting a specific object, it depends on if the object is in the COCO classes, as described below.
\end{itemize}

\small
\begin{lstlisting}[language=json, caption=COCO Classes]
COCO_CLASSES = [
    'person', 'bicycle', 'car', 'motorcycle', 'airplane', 'bus', 'train', 'truck', 
    'boat', 'traffic light', 'fire hydrant', 'stop sign', 'parking meter', 'bench', 
    'bird', 'cat', 'dog', 'horse', 'sheep', 'cow', 'elephant', 'bear', 'zebra', 
    'giraffe', 'backpack', 'umbrella', 'handbag', 'tie', 'suitcase', 'frisbee', 
    'skis', 'snowboard', 'sports ball', 'kite', 'baseball bat', 'baseball glove', 
    'skateboard', 'surfboard', 'tennis racket', 'bottle', 'wine glass', 'cup', 
    'fork', 'knife', 'spoon', 'bowl', 'banana', 'apple', 'sandwich', 'orange', 
    'broccoli', 'carrot', 'hot dog', 'pizza', 'donut', 'cake', 'chair', 'couch', 
    'potted plant', 'bed', 'dining table', 'toilet', 'tv', 'laptop', 'mouse', 
    'remote', 'keyboard', 'cell phone', 'microwave', 'oven', 'toaster', 'sink', 
    'refrigerator', 'book', 'clock', 'vase', 'scissors', 'teddy bear', 'hair drier', 
    'toothbrush'
]
\end{lstlisting}
\normalsize

\begin{itemize}

\item If the object they are looking for is in the COCO classes, still use Yolo11 and COCO. If it is not, use YoloWorld.
For OCR or text extraction, use the Google Cloud Vision API, whose key can be accessed in environment as GOOGLE\_ APPLICATION\_CREDENTIALS
For general LLM tasks not described above, Gemini is available, and its API key can be accessed in environment as GEMINI\_API\_KEY Never use gemini-2.0-flash, if you would use that model, instead use gemini-3-flash-preview.

\item For tools where the custom GPT field is filled to indicate that is what they want, prioritize Gemini live over anything specified above. For these tools, arrange for the specified query to be re-pushed without the user needing to say anything new every 5 seconds.

\item From the GitHub agent, generate all code to be runable within the app, which is described in the ProgramATApp subfolder.

\item From the GitHub agent, all code should be able to use the camera feed provided by the app for visual processing.

\item For any tools that ask for clock face navigation, output should be restricted to 1-3 and 9-12 (others cannot reasonably be captured as they would be behind the camera).

\item Optimize for doing your generation quickly. Extensive documentation after the code is written and tested is not wanted or needed.

\end{itemize}

\section{GitHub Web Interface}
\label{appendix:githubweb}

\begin{figure*}[h!]
    \centering \includegraphics[width=\linewidth,
    alt={Bespoke AT creation process from the GitHub web interface. (1) Open an issue to create the initial prompt. (2) Programming agent processes request by opening a pull request. (3) Test the implemented tool from the mobile app to use the camera. (4) Request iterations by commenting on the open pull request. (5) Once satisfied, merge the pull request to main. Once merged to main branch, modular tool components can be reused/ remixed in other tools.}
    ]{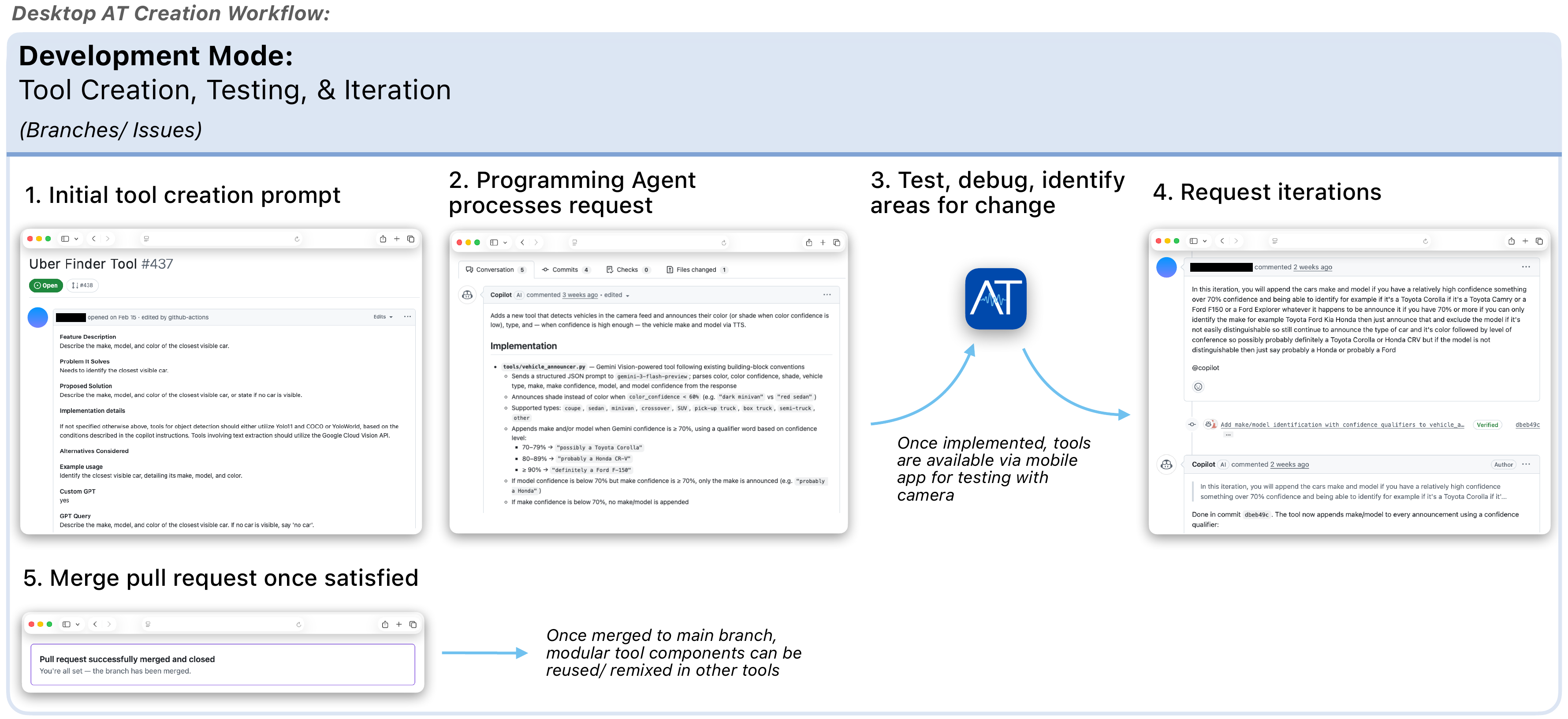}
    \caption{Bespoke AT creation process from the GitHub web interface. (1) Open an issue to create the initial prompt. (2) Programming agent processes request by opening a pull request. (3) Test the implemented tool from the mobile app to use the camera. (4) Request iterations by commenting on the open pull request. (5) Once satisfied, merge the pull request to main.}
    \Description{Bespoke AT creation process from the GitHub web interface. (1) Open an issue to create the initial prompt. (2) Programming agent processes request by opening a pull request. (3) Test the implemented tool from the mobile app to use the camera. (4) Request iterations by commenting on the open pull request. (5) Once satisfied, merge the pull request to main. Once merged to main branch, modular tool components can be reused/ remixed in other tools.}
    \label{fig:web}
\end{figure*}

Users can also create and iterate on bespoke AT from the GitHub web interface, as shown in Figure \ref{fig:web}.

\section{Example prompts and iterations}
\subsection{\GK's hand gesture interpreter}
\subsubsection{Initial prompt}
This prompt was sent from mobile:

I'd like a tool to give me basic scene descriptions and text to speech for videos i'm watching. I want to point my phone at a laptop screen and have the tool speak in real-time what is happening in the video. Keep the descriptions to a minimum and priroritize text that appears on the screen. I don't want the tool to read subtitles. This is for text onscreen that I would not be able to hear from the spoken portion of the video.
\subsubsection{Iteration prompts}
\begin{enumerate}
    \item I changed my mind about a couple things. I want you to make it support live mode like a custom gpt, and i want you to trigger haptic feedback whenever there is something interesting happening visually that I might be missing. So rather than be told basic visual descriptions, I want haptic feedback to play whenever there is information i might be missing without even knowing it, so that i can pause the video and get the basic visual description that way.
    \item The on-text reading does not trigger often enough.
    \item Currently, the tool reads all text aloud on screen. Ignore parts of the screen that are not part of the main video frame. Use edge detection to find the portion of the feed that is the main video and only read aloud text that's part of the video.
    \item Remove all functions and code related to the text to speech portions of the tool. Now i only want the tool to provide haptic feedback when there is something of interest happening onscreen and allowing me to ask the tool what is happening onscreen and get a basic visual description.
    \item Remove the haptic feedback conditional. Just provide basic on screen visual descriptions - real-time.
    \item Restart the whole project. I want a tool that describes basic hand motions that is being used by a person in a video. I want to point my phone at the computer screen and have the tool in real-time describe hand gestures a person is using while speaking.
    \item This tool is working well so far. When describing the hand motions, also include information about the position of the wrist, palm, and arms.
\end{enumerate}

\subsection{\AS's business card reader}
\subsubsection{Initial prompt}
This prompt was sent from the web:

\textbf{Feature Description}

This tool extracts key pieces of information from an image of a card.

Problem It Solves
Blind individuals often need to identify cards whether financial cards like credit card, debit card, etc. or cards for stores, transport, etc.

\textbf{Proposed Solution}

Given an image first detect if it is a card or not.
If it is not a card, inform the user that No Card was Found
If it is a card, check if the text on it is readable, if not readable, give guidance to the user on how to take a better picture i.e. go to better lighting, left part is cutoff so move phone to left, etc.
If text is readable, extract the brand and key details.
Be careful to not output sensitive information i.e. CVV number, SSN, etc.
Decide what is safe to output like last 4 digits of card number, expiry, person's name, etc.

\textbf{Implementation details}

If not specified otherwise above, tools for object detection should either utilize Yolo11 and COCO or YoloWorld, based on the conditions described in the copilot instructions. Tools involving text extraction should utilize the Google Cloud Vision API.

\textbf{Alternatives Considered}

\textbf{Example usage}

User sends an image of a credit card

Output:
\begin{itemize}
    \item Bank of America Card
    \item Card Holder: John Smith
    \item Card Number ends with 1234
\end{itemize}

\textbf{Custom GPT}
No

\textbf{GPT Query}

\textbf{Additional Context}

Unless otherwise specified, in streaming mode, any verbal/text response should be limited to 15 words. No such limit applies to one-shot output.

Write the code for this tool inside the tools folder

\subsubsection{Iteration prompts}
\begin{enumerate}
    \item Tested the tool but it keeps saying No Card Detected, may be along with no card detected message it should describe what it sees instead in 5-10 words for user to know what was captured and can adjust the camera accordingly. @copilot please address this.
    \item Your notes say "Card detection — Uses YoloWorld (yolov8s-world.pt) with custom classes ['credit card', 'debit card', 'card', 'bank card', 'id card'] since none of these are in the COCO dataset. If no card is detected the user is told to hold a card in front of the camera.", that might be the reason for too many no card detected, can we make the tool flexible for lots of card types instead of a fixed set of known categories?
    \item I still don't see scene description when card detection fails. I get "No Card Found ...in front of the cameara" generic message. When this message is reached, scene description should be appended.
    \item Still doesn't work, lets start over by just using the same model as the make up checker tool. Remove the 55 category dataset, custom lighting checking algo, etc. just leverage the vision API with the prompt
    \item I am getting a message "Unable to ... at this time". Can you find bugs and address them?
\end{enumerate}

\subsection{\AZ's kitchen scale reader}
\subsubsection{Initial Prompt}
This prompt was sent from mobile:

You are a helpful, visual assistant with the purpose of reading the current value and measurement on a kitchen scale so it might be in ounces. It might be in grams. It might be in pounds. You will first read the number for example 1 lbs. 1 oz. 1 g or 0.1 pound or whatever it happens to be one detected you will only read the number and you will only read the number followed by the measurement amount, whether that's gram ounce or pound you will not read anything else for example the names of the buttons or anything else if it reads zero you read zero followed by the measurement for example 0 ounces 0 g or 0 pounds.

\subsubsection{Iteration Prompts}
\begin{enumerate}
    \item Add Gemini Live support it should act like a custom GPT so minimize the number of prompting and all that good stuff
    \item Simply have the tool look for a number assume that the number is in grams so read the number followed by the word grams for example 0 g 100 g 1 g 5 g 20 g etc. It should pronounce grams not just say g 
    \item Remove the Gemini life support and redo this app entirely by making it assume that what we're looking at is a scale set to grams so any number you see just append the word grams to it so for example 0 g 10 g 100 g 1000 g so on and so forth make sure to pronounce the word grams instead of just G
\end{enumerate}

\section{Tools created}
\label{appendix:toolsappendix}
Description of all created tools not described in \Cref{tab:tool_table} are available in \Cref{tab:tool_tableappendix}.
\begin{table*}[!b]
    \centering
    \renewcommand{\arraystretch}{1.15}
    \begin{tabular}{p{0.13\linewidth}|l|p{0.26\linewidth}|p{0.2\linewidth}|p{0.1\linewidth}|p{0.09\linewidth}}
        \textbf{Tool} & \textbf{Creator}  &\textbf{ Motivation} & \textbf{Alt. Approach} & \textbf{Mode} & \textbf{App/Web} \\ \hline \hline
        Physical UI Reader & \GK & Get assistance navigating a physical UI, such as buttons on a microwave & Memorize layout of personal appliances & Streaming & App \\ \hline
        Vlog frame analyzer & \GK & Help achieve a well framed, clutter-free, camera shot & Sighted assistance & Streaming & App \\ \hline
        Rubiks cube guidance & \GK & Understand the current face of a Rubiks cube to solve it & Call an Aira agent & Streaming & App \\ \hline
        Cursor finder & \GK*, \KL & Understand the position of a cursor on computer screen, especially if screen reader freezes & Click and see what opens & Live & Web \\ \hline
        Dial reader & \GK & Interpret the position of a dial on an appliance such as an oven & Call sighted assistance (BeMyEyes or Aira) & Take photo & App \\ \hline
        Stain finder & \GK & Identify stain or grime on walls while cleaning & Combination of tactile interaction and sighted assistance tools & Streaming & App \\ \hline
        Washing machine dial reader & \AZ & Interpret the position of a dial on a washing machine & Use sighted assistance or image descriptions & Take photo & App \\ \hline
        Radio screen reader & \AZ & Read the content of a hobbyist radio's digital screen & Use sighted assistance or image descriptions & Streaming & App \\ \hline
        Video describer & \AJ & Describe a video's content including action, subtitles, and content like release date & Use sighted assistance & Streaming & Web \\ \hline
        Object distance for indoor navigation & \AJ & Determine the distance from large physical objects while navigating indoors & Physical tools (e.g. cane) & Streaming & App \\ \hline
        Outfit match validator & \AJ & Determine if an outfit matches and is situationally appropriate for the days events (e.g. casual vs business) & Sighted assistance & Take photo & Web \\ \hline
        Clothes stain checker & \KL & Determine if a clothing item has a stain & Use BeMyAI description & Take photo & App \\ \hline
        Object locator & \KL & Identify common physical objects and help a user navigate to them & Sighted assistance & Streaming & App \\ \hline
        Sock matcher & \AS & Determine if two socks from the laundry match & Ask BeMyAI & Take photo & Web \\
    \end{tabular}
    \vspace{1pc}
    \caption{All tools created in our study not referenced in \Cref{tab:tool_table}. Table includes a description of the tool created, the co-designers' alternative or current approach to accomplish the same task (i.e., what strategy they used without the bespoke AT, if any), the mode of the tool (live, streaming, or photo), and how the tool was created (via the app or web interface). If multiple co-designers made the same tool, details correspond to the co-designer marked with asterisk}
    \label{tab:tool_tableappendix}
\end{table*}

\section{Co-designer demographics}
\label{appendix:demographic_table}

Co-designer age and vision demographics are available in Table \ref{tab:participant_demographic}.

\begin{table*}[!b]
    \renewcommand{\arraystretch}{1}
    \centering
    \begin{tabular}{l|l| l |p{0.55\linewidth} | l}
        \textbf{ID} & \textbf{Age}  & \textbf{\# Phases} &\textbf{Visual acuity} & \textbf{Age of vision loss}\\ \hline \hline
        \GK & 25  & 2 &One eye totally blind, light perception in other & 14\\
        \AZ & 28  & 2 &One eye totally blind, some usable vision in other & birth\\
        \AJ & 42  & 1&Minimal residual vision, some light perception & birth\\
        \KL & 27  & 1 &Minimal residual vision, some light perception & 1\\
        \AS & 33 & 1 &Totally blind & 11\\
    \end{tabular}
    \caption{Co-designer demographics}
    \label{tab:participant_demographic}
\end{table*}

\end{document}